# The NC-CC dichotomy in ruthenium isotopes: Implications for the origin of the late veneer

Jonas Pape [a], Emily A. Worsham [a,b], and Thorsten Kleine [a,c]

[a] Institut für Planetologie, University of Münster, Wilhelm-Klemm-Str. 10, 48149 Münster, Germany

[b] Nuclear and Chemical Sciences Division, Lawrence Livermore National Laboratory, Livermore, CA, USA,

[c] Max Planck Institute for Solar System Research, Justus-von-Liebig-Weg 3, 37077 Göttingen, Germany

Corresponding author: Jonas Pape (jonas.pape@uni-muenster.de)

Accpeted for publication in

*Icarus*

**Abstract**

High-precision mass-independent Ru isotope data for iron meteorites reveal that in three-isotope diagrams including the *p*-process Ru nuclides $^{96}$Ru or $^{98}$Ru, non-carbonaceous (NC) meteorites plot along an *s*-process mixing line, while the carbonaceous (CC) type irons plot as a cluster off the NC-line. As previously observed for Mo, this offset can be accounted for by an *r*-process excess in CC over NC materials. As a highly siderophile element, Ru in the bulk silicate Earth (BSE) is thought to predominantly derive from the late veneer, making the Ru isotope dichotomy a powerful genetic tracer of Earth's late accretionary epoch. The BSE and the isotopically anomalous Itsaq Gneiss Complex as defined in prior studies plot on the NC-line defined in this study, suggesting a predominantly NC late veneer that may have evolved over time from more *s*-process-enriched to more *s*-process-depleted compositions. However, more precise $^{96}$Ru and $^{98}$Ru measurements of the BSE's composition are needed to more reliably determine the genetic heritage of the late veneer and to quantify as to whether it contains CC material.

## 1. Introduction

Nucleosynthetic isotope anomalies among meteorites arise through the heterogeneous distribution of presolar components in the protoplanetary disk (e.g., Dauphas and Schauble, 2016). These anomalies have been identified for a large number of elements (e.g., Ti, Cr, Fe, Ni, Mo, Ru) and distinguish between non-carbonaceous (NC) and carbonaceous (CC) meteorites (Trinquier et al., 2007; Warren, 2011; Budde et al, 2016), which are presumed to represent the inner and outer solar system, respectively (Warren, 2011; Kruijer et al., 2017). These nucleosynthetic anomalies have been used in a wide range of studies focusing, for instance, on the genetic heritage of meteorites and their components, the early dynamical evolution of the solar accretion disk, and the provenance of Earth's building materials (e.g., Dauphas et al., 2004; Qin and Carlson, 2016; Dauphas 2017; Poole et al., 2017; Bermingham et al., 2020; Ek et al., 2020; Kleine et al., 2020; Spitzer et al., 2022). The isotope anomalies in Mo have proven particularly useful because in plots including *p*-process Mo isotopes (i.e. $^{92}$Mo or $^{94}$Mo), meteorites plot on two approximately parallel lines, where the isotope variations along each line predominantly reflect *s*-process heterogeneity, while the offset between the two lines reflects an *r*-process excess in CC over NC materials (Budde et al., 2016; Kruijer et al., 2017; Poole et al., 2017; Worsham et al., 2017; Spitzer et al., 2020). Moreover, as a moderately siderophile element, the Mo in the bulk silicate Earth (BSE) predominantly derives from the last ~10-20% of Earth's growth (Dauphas, 2017), regardless of whether the Earth accreted heterogeneously (e.g., Dauphas et al., 2017; Budde et al., 2019; Burkhardt et al., 2021; Dauphas et al., 2024) or homogeneously (Sossi and Bower, 2026). As such, the BSE's Mo isotope composition provides constraints on the genetic heritage of Earth's late-stage building blocks (Budde et al., 2018; Bermingham et al., 2025).

Like Mo, the highly siderophile element (HSE) Ru has potential as a strong, single-element discriminator between NC and CC materials. Ruthenium has seven stable isotopes with variable contributions from *p*-, *s*-, and *r*-process nucleosynthesis: $^{96}$Ru (*p*-only), $^{98}$Ru (*p*-only), $^{99}$Ru (31.2% *s*-, 68.8% *r*-process), $^{100}$Ru (*s*-only), $^{101}$Ru (14.1% *s*-, 85.9% *r*-process ), $^{102}$Ru (47.6% *s*-, 52.4% *r*-process ), and $^{104}$Ru (*r*-only) (Prantzos et al., 2020). As such, Ru isotopes should also allow distinguishing between *s*- and *r*-process variations, which would allow resolving the NC-CC dichotomy solely by plotting two Ru isotope ratios against each other. As a HSE, the Ru in the present-day BSE is thought to predominantly derive from late accretion (or late veneer), which is defined as the final ~0.5% of Earth's accretion following the cessation of core formation (e.g., Walker, 2009). Evidence for late accretion comes primarily from the elevated abundances and approximately chondritic relative proportions of the HSEs in the BSE, which are unlikely to have been established by metal-silicate equilibration during core formation (e.g., Chou et al., 1978; Becker et al., 2006). Metal-silicate partition experiments indicate that although the HSEs become less siderophile with increasing pressure and temperature (e.g., Righter et al., 2008), their abundances would nevertheless be fractionated from each other (e.g., Brenan and McDonough, 2009). Thus, while some of the HSEs in Earth's mantle may have been retained in the mantle after the last core formation event, to account for their approximately chondritic relative abundances, most of the HSEs in the BSE were likely added by late accretion (e.g., Brenan and McDonough, 2009). The late-accreted mass on Earth may, therefore, have been lower than the aforementioned ~0.5%, which is derived by assuming that the BSE's HSEs solely come from late accretion. A lower late-accreted mass has also been inferred based on the BSE's abundances of S, Se, and Te (Calvo et al., 2026). In the specific case of Ru it has been proposed that about 20% of the BSE's Ru derives from prior to late accretion, while the rest has been added by the late veneer (Rubie et al., 2016). This makes the nucleosynthetic isotope signatures of Ru particularly useful for determining the genetics of the late veneer (Fischer-Gödde and Kleine, 2017; Bermingham and Walker, 2017).

Until now, Ru isotope studies have focused mainly on isotope anomalies in $^{100}$Ru, and data for the lower abundance *p*-process nuclides $^{96}$Ru and $^{98}$Ru were typically not precise enough to resolve the expected NC-CC dichotomy for Ru (Chen et al., 2010; Fischer-Gödde et al., 2015; Fischer-Gödde and Kleine, 2017; Bermingham and Walker, 2017; Bermingham et al., 2018; Hopp et al., 2018). Nevertheless, the $^{100}$Ru data for especially iron meteorites reveal distinct isotopic compositions for NC and CC bodies, where the NC irons exhibit variable $^{100}$Ru deficits compared to the BSE, while the CC irons are characterized by an approximately constant and larger $^{100}$Ru deficit (e.g., Fischer-Gödde et al., 2015; Worsham et al., 2019; Tornabene et al., 2020). Carbonaceous chondrites exhibit more variable $^{100}$Ru compositions, which probably reflect the redistribution of isotopically anomalous Ru during processes on the chondrite parent bodies, and as such may not reflect true bulk parent body compositions (Fischer-Gödde et al., 2017). Since the BSE's $^{100}$Ru composition overlaps with those of NC meteorites, most studies concluded that the BSE's Ru is purely NC, implying an NC-dominated late veneer (e.g., Fischer-Gödde and Kleine, 2017; Bermingham and Walker, 2017; Bermingham et al., 2018; Worsham and Kleine, 2021). However, based on $^{100}$Ru excesses measured for samples from the Itsaq Gneiss Complex, Fischer-Gödde et al. (2020) argued that the BSE's Ru derives from a mixture of isotopically more anomalous NC with CI chondrite-like materials, implying a CC-dominated late veneer. One way of distinguishing between these disparate interpretations is to also consider $^{96}$Ru and $^{98}$Ru because for these two nuclides, *s*- and *r*-process variations result in distinct isotopic patterns. Consequently, the characteristic *r*-process excess of CC over NC materials is expected to result in distinct $^{96}$Ru and $^{98}$Ru isotope compositions between NC and CC. Thus, establishing whether a NC-CC dichotomy can be resolved when two Ru isotope ratios are used together would provide a new, powerful tool for evaluating the genetic heritage of the late veneer.

We optimized our analytical techniques for $^{96}$Ru and $^{98}$Ru measurements and report new high-precision Ru isotope data for fourteen NC- and CC-type iron meteorites. These data are used to discuss analytical challenges inherent in acquiring precise and accurate $^{96}$Ru and $^{98}$Ru data, to demonstrate that the NC-CC dichotomy can be resolved in Ru three-isotope plots, and to evaluate the contribution of NC and CC materials to the late veneer.

## 2. Samples and Methods

### *2.1 Samples and sample preparation*

Fourteen iron meteorites were selected for this study, including both NC (IAB, IC, IIAB, IIIAB, IVA) and CC (IIC, IID, IIIF, Chinga) irons (Table 1). Most of these samples were previously analyzed for their Pt isotope compositions to monitor cosmic-ray exposure (CRE) effects, and no significant Pt isotope anomalies are present in these samples (Kruijer et al., 2013, 2017, Worsham et al., 2019; Spitzer et al., 2020). We additionally measured the Pt isotopic composition for some samples from aliquots of the digestion solutions used for Ru isotope analyses. With the exception of Henbury and Unter-Mässing ($\varepsilon^{196}$Pt = 0.15 ± 0.07 and 0.12 ± 0.02, respectively) all samples of this study show no resolvable Pt isotope anomalies (Table 1). Even CRE effects of the size measured for Henbury and Unter-Mässing only result in a $\varepsilon^{100}$Ru correction of approximately 2-3 ppm, which is within uncertainty of the data reported here. Finally, no Pt isotope data exist for the ungrouped iron Chinga, but Os isotope data indicate that this meteorite was not or only minimally affected by CRE (Bermingham et al., 2018; Yokoyama et al., 2019; Corrigan et al., 2022). Consequently, no CRE corrections were required for any of the data reported here.

TABLE 1. Measured Ru isotope data and corrected group means for iron meteorites.

| Sample | N | Normalized to $^{99}$Ru/$^{101}$Ru | | | | | Normalized to $^{102}$Ru/$^{100}$Ru | | | | | |
|---|---|---|---|---|---|---|---|---|---|---|---|---|
| | | ε$^{96}$Ru (± 2σ) | ε$^{98}$Ru (± 2σ) | ε$^{100}$Ru (± 2σ) | ε$^{102}$Ru (± 2σ) | ε$^{104}$Ru (± 2σ) | ε$^{96}$Ru (± 2σ) | ε$^{98}$Ru (± 2σ) | ε$^{99}$Ru (± 2σ) | ε$^{101}$Ru (± 2σ) | ε$^{104}$Ru (± 2σ) | ε$^{196}$Pt (± 2σ) |
| ***Non-carbonaceous*** | | | | | | | | | | | | |
| **IAB-MG** | | | | | | | | | | | | |
| Canyon Diablo 1) | 5 | 0.34 ± 0.09 | 0.28 ± 0.08 | -0.08 ± 0.03 | -0.01 ± 0.02 | | 0.60 ± 0.04 | 0.45 ± 0.16 | 0.13 ± 0.05 | 0.05 ± 0.01 | | 0.01 ± 0.07 |
| Canyon Diablo replicate 2) | 4 | 0.27 ± 0.10 | 0.37 ± 0.09 | -0.05 ± 0.04 | -0.04 ± 0.03 | 0.12 ± 0.17 (2)2) | 0.39 ± 0.10 | 0.46 ± 0.09 | 0.07 ± 0.06 | 0.05 ± 0.04 | 0.11 ± 0.10 (2) 2) | |
| Campo del Cielo 1) | 7 | 0.53 ± 0.03 | 0.40 ± 0.07 | -0.12 ± 0.01 | -0.03 ± 0.03 | | 0.82 ± 0.09 | 0.61 ± 0.07 | 0.16 ± 0.03 | 0.08 ± 0.02 | | 0.00 ± 0.07 |
| Campo del Cielo replicate 1) | 5 | 0.42 ± 0.12 | 0.52 ± 0.16 | -0.07 ± 0.03 | -0.01 ± 0.04 | 0.18 ± 0.37 (4)1) | 0.64 ± 0.07 | 0.69 ± 0.13 | 0.11 ± 0.04 | 0.04 ± 0.03 | 0.15 ± 0.15 (4) 1) | |
| **IAB-MG mean** | | **0.40 ± 0.06** | **0.40 ± 0.05** | **-0.08 ± 0.01** | **-0.02 ± 0.01** | **0.16 ± 0.19** | **0.64 ± 0.08** | **0.56 ± 0.06** | **0.12 ± 0.02** | **0.06 ± 0.01** | **0.13 ± 0.08** | |
| **IAB-MG mean corrected 4)** | | **0.15 ± 0.09** | **0.36 ± 0.09** | **-0.06 ± 0.02** | **-0.07 ± 0.02** | **-0.04 ± 0.20** | **0.34 ± 0.14** | **0.45 ± 0.09** | **0.07 ± 0.04** | **0.08 ± 0.02** | **0.02 ± 0.09** | |
| **IAB-sLL** | | | | | | | | | | | | |
| Toluca 1) | 6 | 0.75 ± 0.07 | 0.37 ± 0.06 | -0.08 ± 0.03 | 0.07 ± 0.05 | | 1.10 ± 0.18 | 0.60 ± 0.16 | 0.15 ± 0.06 | 0.00 ± 0.02 | | 0.03 ± 0.07 |
| Toluca replicate 2) | 5 | 0.50 ± 0.12 | 0.35 ± 0.21 | -0.06 ± 0.06 | 0.01 ± 0.05 | 0.29 ± 0.17 (3)2) | 0.67 ± 0.20 | 0.48 ± 0.20 | 0.09 ± 0.10 | 0.03 ± 0.05 | 0.22 ± 0.10 (3) 2) | |
| **IAB-sLL mean** | | **0.64 ± 0.10** | **0.36 ± 0.08** | **-0.07 ± 0.02** | **0.04 ± 0.04** | **0.29 ± 0.17** | **0.90 ± 0.19** | **0.55 ± 0.11** | **0.12 ± 0.05** | **0.01 ± 0.02** | **0.22 ± 0.10** | |
| **IAB-sLL mean corrected 4)** | | **0.39 ± 0.12** | **0.32 ± 0.11** | **-0.05 ± 0.03** | **-0.01 ± 0.05** | **0.09 ± 0.18** | **0.60 ± 0.22** | **0.44 ± 0.13** | **0.07 ± 0.06** | **0.03 ± 0.03** | **0.11 ± 0.11** | |
| **IC** | | | | | | | | | | | | |
| Chihuahua City 1) | 4 | 0.55 ± 0.18 | 0.40 ± 0.04 | -0.37 ± 0.01 | -0.03 ± 0.21 | 0.38 ± 0.31 (3)1) | 1.64 ± 0.54 | 1.12 ± 0.28 | 0.56 ± 0.13 | 0.20 ± 0.11 | 0.29 ± 0.34 (3) 1) | 0.07 ± 0.07 5) |
| Mount Dooling 1) | 7 | 0.56 ± 0.07 | 0.51 ± 0.06 | -0.41 ± 0.04 | -0.17 ± 0.03 | | 1.44 ± 0.09 | 1.15 ± 0.07 | 0.53 ± 0.04 | 0.29 ± 0.02 | | -0.01 ± 0.04 6) |
| Mount Dooling replicate 1) | 8 | 0.50 ± 0.05 | 0.38 ± 0.07 | -0.41 ± 0.01 | -0.16 ± 0.04 | | 1.44 ± 0.07 | 1.09 ± 0.12 | 0.53 ± 0.03 | 0.29 ± 0.02 | | |
| **IC mean** | | **0.53 ± 0.04** | **0.41 ± 0.06** | **-0.40 ± 0.01** | **-0.14 ± 0.04** | **0.38 ± 0.31** | **1.48 ± 0.09** | **1.11 ± 0.06** | **0.53 ± 0.02** | **0.27 ± 0.02** | **0.29 ± 0.34** | |
| **IC mean corrected 4)** | | **0.28 ± 0.08** | **0.37 ± 0.09** | **-0.38 ± 0.02** | **-0.19 ± 0.05** | **0.18 ± 0.31** | **1.18 ± 0.14** | **1.00 ± 0.09** | **0.48 ± 0.04** | **0.29 ± 0.03** | **0.18 ± 0.34** | |
| **IIAB** | | | | | | | | | | | | |
| North Chile 2) | 4 | 0.48 ± 0.13 | 0.53 ± 0.29 | -0.41 ± 0.04 | -0.15 ± 0.09 | 0.20 ± 0.17 (2)2) | 1.44 ± 0.15 | 1.21 ± 0.23 | 0.54 ± 0.02 | 0.28 ± 0.06 | 0.07 ± 0.10 (2) 2) | 0.02 ± 0.03 6) |
| **IIAB mean** | | **0.48 ± 0.13** | **0.53 ± 0.29** | **-0.41 ± 0.04** | **-0.15 ± 0.09** | **0.20 ± 0.17** | **1.44 ± 0.15** | **1.21 ± 0.23** | **0.54 ± 0.02** | **0.28 ± 0.06** | **0.07 ± 0.10** | |
| **IIAB mean corrected 4)** | | **0.23 ± 0.14** | **0.49 ± 0.30** | **-0.39 ± 0.05** | **-0.20 ± 0.09** | **0.00 ± 0.18** | **1.14 ± 0.19** | **1.10 ± 0.24** | **0.49 ± 0.04** | **0.30 ± 0.06** | **-0.04 ± 0.11** | |
| **IIIAB** | | | | | | | | | | | | |
| Cape York 1) | 7 | 0.65 ± 0.08 | 0.54 ± 0.08 | -0.64 ± 0.03 | -0.23 ± 0.04 | | 2.06 ± 0.14 | 1.53 ± 0.11 | 0.83 ± 0.05 | 0.44 ± 0.03 | | 0.01 ± 0.07 7) |
| Cape York replicate 3) | 2 | 0.90 ± 0.25 | 0.58 ± 0.19 | -0.69 ± 0.06 | -0.18 ± 0.05 | 0.59 ± 0.17 | 2.61 ± 0.40 | 1.76 ± 0.20 | 0.95 ± 0.09 | 0.42 ± 0.06 | 0.26 ± 0.10 | |
| Henbury 3) | 6 | 0.59 ± 0.04 | 0.54 ± 0.10 | -0.54 ± 0.03 | -0.19 ± 0.03 | 0.23 ± 0.06 | 1.84 ± 0.12 | 1.46 ± 0.09 | 0.72 ± 0.04 | 0.37 ± 0.02 | 0.08 ± 0.03 | 0.15 ± 0.07 7) |
| Henbury replicate 1 3) | 4 | 1.05 ± 0.04 | 0.62 ± 0.27 | -0.60 ± 0.04 | -0.16 ± 0.04 | 0.65 ± 0.08 | 2.58 ± 0.15 | 1.65 ± 0.25 | 0.84 ± 0.04 | 0.38 ± 0.03 | 0.32 ± 0.04 | |
| Henbury replicate 2 3) | 5 | 0.70 ± 0.11 | 0.55 ± 0.21 | -0.56 ± 0.04 | -0.18 ± 0.04 | 0.35 ± 0.08 | 2.13 ± 0.09 | 1.47 ± 0.20 | 0.77 ± 0.03 | 0.37 ± 0.04 | 0.15 ± 0.06 | |
| **IIIAB mean** | | **0.73 ± 0.07** | **0.56 ± 0.05** | **-0.60 ± 0.02** | **-0.19 ± 0.02** | **0.41 ± 0.10** | **2.15 ± 0.13** | **1.54 ± 0.06** | **0.80 ± 0.03** | **0.39 ± 0.02** | **0.18 ± 0.05** | |
| **IIIAB mean corrected 4)** | | **0.48 ± 0.08** | **0.52 ± 0.09** | **-0.58 ± 0.03** | **-0.24 ± 0.03** | **0.21 ± 0.11** | **1.85 ± 0.17** | **1.43 ± 0.09** | **0.75 ± 0.04** | **0.41 ± 0.03** | **0.07 ± 0.06** | |
| **IVA** | | | | | | | | | | | | |
| Muonionalusta 2) | 6 | 0.54 ± 0.03 | 0.49 ± 0.14 | -0.29 ± 0.03 | -0.06 ± 0.03 | 0.32 ± 0.17 (2)2) | 1.31 ± 0.11 | 0.98 ± 0.13 | 0.40 ± 0.04 | 0.17 ± 0.03 | 0.18 ± 0.10 (2)2) | -0.11 ± 0.07 |
| Muonionalusta replicate 3) | 4 | 0.22 ± 0.15 | 0.28 ± 0.15 | -0.27 ± 0.08 | -0.13 ± 0.05 | 0.04 ± 0.10 | 0.81 ± 0.14 | 0.69 ± 0.02 | 0.35 ± 0.13 | 0.21 ± 0.03 | 0.03 ± 0.07 | |
| **IVA mean** | | **0.41 ± 0.13** | **0.41 ± 0.11** | **-0.28 ± 0.03** | **-0.09 ± 0.03** | **0.13 ± 0.16** | **1.11 ± 0.20** | **0.86 ± 0.13** | **0.38 ± 0.04** | **0.18 ± 0.02** | **0.08 ± 0.09** | |
| **IVA mean corrected 4)** | | **0.16 ± 0.14** | **0.37 ± 0.13** | **-0.26 ± 0.04** | **-0.14 ± 0.04** | **-0.07 ± 0.17** | **0.81 ± 0.23** | **0.75 ± 0.15** | **0.33 ± 0.05** | **0.20 ± 0.03** | **-0.03 ± 0.10** | |
| ***Carbonaceous*** | | | | | | | | | | | | |
| **IIC** | | | | | | | | | | | | |
| Unter-Mässing [1)] | 5 | 0.48 ± 0.06 | 0.30 ± 0.05 | -0.97 ± 0.02 | -0.38 ± 0.04 | | 2.68 ± 0.08 | 1.85 ± 0.03 | 1.28 ± 0.01 | 0.68 ± 0.04 | | 0.12 ± 0.02[8)] |
| Unter-Mässing replcicate [1)] | 4 | 0.22 ± 0.13 | 0.22 ± 0.11 | -0.95 ± 0.04 | -0.39 ± 0.09 | -0.10 ± 0.10 (7)[1)] | 2.31 ± 0.02 | 1.71 ± 0.10 | 1.23 ± 0.06 | 0.67 ± 0.06 | -0.02 ± 0.34 (7)[1)] | |
| **IIC mean** | | **0.37 ± 0.12** | **0.26 ± 0.05** | **-0.96 ± 0.02** | **-0.38 ± 0.03** | **-0.10 ± 0.10** | **2.51 ± 0.15** | **1.79 ± 0.07** | **1.25 ± 0.03** | **0.67 ± 0.02** | **-0.02 ± 0.34** | |
| **IIC mean corrected [4)]** | | **0.12 ± 0.13** | **0.22 ± 0.09** | **-0.94 ± 0.03** | **-0.43 ± 0.04** | **-0.30 ± 0.11** | **2.21 ± 0.19** | **1.68 ± 0.10** | **1.20 ± 0.04** | **0.69 ± 0.03** | **-0.13 ± 0.34** | |
| **IID** | | | | | | | | | | | | |
| Bridgewater [1)] | 8 | 0.86 ± 0.03 | 0.32 ± 0.07 | -1.03 ± 0.02 | -0.31 ± 0.03 | | 3.38 ± 0.10 | 2.08 ± 0.09 | 1.41 ± 0.04 | 0.68 ± 0.01 | | -0.01 ± 0.02[5)] |
| Bridgewater replicate [2)] | 6 | 0.12 ± 0.08 | 0.19 ± 0.17 | -1.01 ± 0.03 | -0.42 ± 0.03 | -0.07 ± 0.17 (2)[2)] | 2.36 ± 0.07 | 1.81 ± 0.15 | 1.30 ± 0.05 | 0.72 ± 0.02 | -0.22 ± 0.10 (2)[2)] | |
| Rodeo [1)] | 2 | 0.56 ± 0.25 | 0.39 ± 0.19 | -1.01 ± 0.06 | -0.38 ± 0.05 | 0.27 ± 0.29 (4)[1)] | 2.84 ± 0.40 | 2.04 ± 0.20 | 1.33 ± 0.09 | 0.69 ± 0.06 | -0.19 ± 0.09 (4)[1)] | 0.00 ± 0.11[7)] |
| **IID mean** | | **0.54 ± 0.19** | **0.28 ± 0.07** | **-1.02 ± 0.02** | **-0.36 ± 0.03** | **0.16 ± 0.06** | **2.93 ± 0.27** | **1.98 ± 0.09** | **1.36 ± 0.04** | **0.69 ± 0.01** | **-0.20 ± 0.05** | |
| **IID mean corrected [4)]** | | **0.29 ± 0.20** | **0.24 ± 0.10** | **-1.00 ± 0.03** | **-0.41 ± 0.04** | **-0.04 ± 0.07** | **2.63 ± 0.29** | **1.87 ± 0.11** | **1.31 ± 0.05** | **0.71 ± 0.02** | **-0.31 ± 0.06** | |
| **IIIF** | | | | | | | | | | | | |
| Clark County [1)] | 7 | 0.54 ± 0.08 | 0.28 ± 0.07 | -0.95 ± 0.02 | -0.33 ± 0.07 | | 2.77 ± 0.13 | 1.86 ± 0.08 | 1.27 ± 0.04 | 0.64 ± 0.04 | | 0.07 ± 0.05[9)] |
| **IIIF mean** | | **0.54 ± 0.08** | **0.28 ± 0.07** | **-0.95 ± 0.02** | **-0.33 ± 0.07** | | **2.77 ± 0.13** | **1.86 ± 0.08** | **1.27 ± 0.04** | **0.64 ± 0.04** | | |
| **IIIF mean corrected [4)]** | | **0.29 ± 0.11** | **0.24 ± 0.10** | **-0.93 ± 0.03** | **-0.38 ± 0.08** | | **2.47 ± 0.17** | **1.75 ± 0.11** | **1.22 ± 0.05** | **0.66 ± 0.05** | | |
| **Ungrouped** | | | | | | | | | | | | |
| Chinga [1)] | 7 | 0.53 ± 0.06 | 0.27 ± 0.07 | -1.06 ± 0.02 | -0.40 ± 0.02 | | 2.97 ± 0.05 | 2.03 ± 0.12 | 1.41 ± 0.03 | 0.73 ± 0.02 | | |
| Chinga corrected [4)] | | 0.28 ± 0.09 | 0.23 ± 0.10 | -1.04 ± 0.03 | -0.45 ± 0.03 | | 2.67 ± 0.12 | 1.92 ± 0.14 | 1.36 ± 0.04 | 0.75 ± 0.03 | | |

Uncertainties represent the 95% confidence intervals of the mean (i.e., (s.d. ×t0.95, N-1)/√N) for N ≥ 4, and for N < 4 the 2 s.d. of the repeated analyses of the reference material NIST 129c.

[1)] Measured applying cup config. #1 and Ni H cones. The $\varepsilon^{104}$Ru values (where reported) were determined from measurements using the same sample solution but cup config. #2 and at a concentration of 100 ppb Ru; the number in brackets denotes number of analyses (see Methods for more details).

[2)] Measured applying cup configurations #1 and #2, and cone setups #1 and #2 using the same sample solution (see *Methods* for more details). The number in brackets for $\varepsilon^{104}$Ru values denotes number of analyses applying cup config. #2.

[3)] Measured applying cup config. #2 and cone setup #2 (see *Methods* for more details).

[4)] Group mean values corrected for a small analytical artefact. Values used for correction are 0.25±0.05, 0.04±0.04, -0.02±0.01, 0.05±0.01, and 0.20±0.03 for $\varepsilon^{96}$Ru, $\varepsilon^{98}$Ru, $\varepsilon^{100}$Ru, $\varepsilon^{102}$Ru, and $\varepsilon^{104}$Ru, respectively, (normalized to $^{99}$Ru/$^{101}$Ru), and 0.30±0.09, 0.11±0.04, 0.05±0.02, -0.02±0.01, and 0.11±0.02 for $\varepsilon^{96}$Ru, $\varepsilon^{98}$Ru, $\varepsilon^{99}$Ru, $\varepsilon^{101}$Ru, and $\varepsilon^{104}$Ru, respectively, (normalized to $^{102}$Ru/$^{100}$Ru) and were calculated as the weighted mean of the unaccounted-for fractionation offsets from zero observed in repeated analyses of Ru-doped NIST standard and Ru-doped Muonionalusta iron meteorite (see text for more details).

[5)] Kruijer et al., 2017.

[6)] Spitzer et al., 2020.

[7)] Fischer-Gödde et al., 2015.

[8)] Worsham et al., 2019.

[9)] Pape et al., 2023.

[9)] Replicates from the same sample digestion solution but with seperate Ru chemistry.

To verify our methods for $^{96}$Ru and $^{98}$Ru isotope measurements, we repeatedly processed aliquots of the NIST SRM 129c steel doped with the Alfa Aesar Specpure™ Ru analytical standard (LOT 61300952) through the full analytical procedure (Table S1). For each experiment, ca. 0.4 mL of 10 ppm Ru solution standard were added to 0.2 g of NIST SRM 129c metal chips dissolved in 5 ml of double-distilled 6 M HCl. Since this steel does not contain any detectable Ru (based on NIST's Certificate of Analysis), these measurements are expected to yield the Ru isotope composition of the Alfa Aesar standard. As a further analytical test, we doped aliquots of the IVA iron meteorite Muonionalusta with different amounts of terrestrial Ru from the same standard solution (Table S2). This meteorite was chosen because of its comparatively low Ru content (for an iron meteorite) and because a relatively large amount of this sample was available for analysis. The digestion of a single large piece (~2.5 g) of Muonionalusta was split into four aliquots, three of which (#2-#4) were doped with different amounts of terrestrial Ru (Alfa Aesar standard), while one aliquot did not contain any terrestrial Ru (#1).

Iron meteorite samples were cut from larger pieces, cleaned using SiC abrasive, and rinsed with ethanol in an ultrasonic bath. Meteorite chunks of up to a few grams (typically ~0.5 – 2 g, depending on the Ru concentration) were weighed into 60 ml Savillex PFA beakers and leached in 6 M HCl for 15 minutes at 70 C°. The leachate was removed and the samples were repeatedly rinsed with Milli-Q water before digestion.

*2.2 Ru purification and mass spectrometry*

The methods for Ru purification by ion exchnage chromatography and micro-distillation closely followed those described in Fischer-Gödde et al. (2015) and Worsham et al. (2019) and are described in detail in the Supplementary Information. In brief, following digestion in 20-30 ml double-distilled 6 M HCl at 120 °C, the samples were treated with reverse *aqua regia* and then HCl, re-dissolved in 0.2 M HCl and loaded onto 12 ml BioRad Econo-Columns® filled with 10 ml BioRad AG 50W-X8 resin (100-200 mesh). On this column, Ru together with other HSEs was separated from the sample matrix. A maximum of 0.3 g of iron meteorite was loaded onto single columns and larger samples were split over several columns and recombined prior to the subsequent micro-distillation. The Ru (+HSE) cuts were dried at ~80 °C, re-dissolved in three drops of reverse *aqua regia*, evaporated to near dryness, diluted with Milli-Q water and transferred to the lid of conical 5 ml Savillex beakers, and dried at ~80 °C for micro-distillation using 50 µl di-chromate ($H_2SO_4$-$CrO_3$) solution (Birck et al., 1997; Fischer-Gödde et al., 2015). During the distillation, Ru was evaporated into a droplet of concentrated HBr, which was then dried and taken up in 0.28M $HNO_3$ for isotope analysis. As in prior studies (e.g., Fischer-Gödde et al., 2015; Bermingham et al., 2016; Fischer-Gödde et al., 2024), Ru yields for the cation columns were typically >95%, while those of the micro-distillation varied between ~40 and >90 %. Notably, the yields for meteorite samples were generally higher compared to Ru-doped NIST steels, and significantly higher than for the pure Alfa Aesar Specpure™ standard, suggesting that differences in the amount and type of sample exert some control on the overall yield of the micro-distillation.

The Ru isotope measurements were performed on a Thermo Scientific Neptune *Plus* MC-ICP-MS at the Institut für Planetologie in Münster. The instrument setup was similar to our previous Ru isotope studies (Fischer-Gödde et al., 2015; Hopp et al., 2018; Worsham et al., 2019; Worsham and Kleine, 2021). The sample solutions were introduced into the mass spectrometer using a Savillex C-flow nebulizer (40-50 µL $min^{-1}$) attached to a CETAC Aridus2 (or Aridus3) desolvator. To achieve higher ion beam intensities on $^{96}$Ru and $^{98}$Ru, most sample solutions were measured at a concentration of 500 ppb Ru (compared to 100 ppb Ru typically used in prior Ru MC-ICP-MS studies), resulting in total ion beam intensities of $\sim 6 \times 10^{-10}$ and $\sim 1 \times 10^{-9}$ A. All meteorites were analyzed using conventional Ni H-cones (cone-setup-#1), but some sample solutions (and the IIIAB iron Henbury) were also re-measured using a combination of Ni X-skimmer and H-sampler (cone-setup-#2). The latter cone-setup significantly increases sensitivity, potentially allowing for higher precision for the two low-abundance isotopes $^{96}$Ru and $^{98}$Ru, but also increases the risk of more substantial molecular interferences, particularly of $^{40}Ar_2{}^{16}O$ and $^{56}Fe^{40}Ar$ on $^{96}$Ru. However, we observed no systematic differences among the results for samples measured using both cone-setups (Table 1 and Table S3). Each Ru isotope analysis comprised an on-peak background measurement of a solution blank for 40 cycles of 8.4 s each, followed by a sample or standard measurement of 100 cycles of 8.4 s each. Measurements of unknowns were bracketed with measurement of the Alfa Aesar standard matching the concentrations of the samples to better than ±5%. The data are internally normalized to either $^{99}Ru/^{101}Ru = 0.7450754$ or $^{102}Ru/^{100}Ru = 2.51449$ (Chen et al., 2010) using the exponential law. While prior studies used the first of these normalization (e.g., Chen et al., 2010, Fischer-Gödde et al., 2015, Worsham et al., 2019), when focusing on $^{96}$Ru and $^{98}$Ru, we have found that the $^{102}Ru/^{100}Ru$ normalization is also useful, because the nucleosynthetic Ru isotope variations are generally larger and result in more strongly correlated isotope variations (Fig. S1). The data for the two normalizations are reported as $\varepsilon^{i}Ru_{99/101}$ and $\varepsilon^{i}Ru_{102/100}$ values as the parts-per-10,000 deviations from the terrestrial standard as follows:

$$\varepsilon^{i}Ru_{99/101} = [(^{i}Ru/^{101}Ru)_{sample} / {}^{i}Ru/^{101}Ru)_{standard} - 1] \times 10{,}000$$

and

$$\varepsilon^{i}Ru_{102/100} = [(^{i}Ru/^{100}Ru)_{sample} / {}^{i}Ru/^{100}Ru)_{standard} - 1] \times 10{,}000$$

Potential isobaric interferences of Zr, Mo, and Pd on $^{96}Ru$, $^{98}Ru$,$^{102}Ru$, and $^{104}Ru$ were monitored by measuring the ion beams on $^{91}Zr$, $^{97}Mo$, and $^{105}Pd$. The cup setup of the Neptune *Plus* did not allow measuring all these isotopes simultaneously during a single run, and to circumvent this problem, we have analyzed most of the samples using two different cup-configurations. In cup-configuration #1, $^{104}Ru$ was not measured and the potential interference of $^{102}Pd$ on $^{102}Ru$ was not monitored, while in cup-configuration #2, the potential interference of $^{96}Zr$ on $^{96}Ru$ could not be monitored. However, the Mo/Ru, Zr/Ru, and Pd/Ru ratios of all samples were determined prior to isotope analyses from small aliquots of the purified sample solutions. Most samples had Pd/Ru and Zr/Ru ratios $<10^{-6}$, which required no interference correction. Only Rodeo had a higher Pd/Ru of $3 \times 10^{-4}$, requiring an interference corrections of ~1.7 ε-units on $^{104}Ru$. Most samples had Zr/Ru ratios between $10^{-6}$ and $10^{-7}$ resulting in insignificant Zr interference corrections of ~0.01–0.03 $\varepsilon^{96}Ru$ (both normalizations). A few samples had Zr/Ru of up to $10^{-5}$, resulting in interference corrections of ~0.1–0.2 $\varepsilon^{96}Ru$ but tests using Zr-doped Ru standard solution showed that Zr/Ru ratios of up to $8 \times 10^{-4}$ can be corrected for accurately (Fig. S2). The Mo/Ru ratios of the samples were $< 5 \times 10^{-5}$ and tests using Mo-doped Ru standard solution showed that interference corrections are accurate for Mo/Ru ratios up to $2 \times 10^{-4}$ (Fig. S2).

## 3. Results

### *3.1 Ru-doped NIST SRM 129c and Muonionalusta*

The Ru isotope data for the Ru-doped NIST SRM 129c samples are reported in Table S1. Most of these samples have positive $\varepsilon^{96}Ru$ and $\varepsilon^{104}Ru$ values, using both normalizations. Additionally, small positive anomalies were observed for $\varepsilon^{102}Ru_{99/101}$ and $\varepsilon^{98}Ru_{102/100}$, as well as small negative anomalies for $\varepsilon^{101}Ru_{102/100}$. Overall, these systematics result in a U-shaped isotope pattern for the Ru-doped NIST SRM 129c measurements (Fig. S3).

The Ru isotope data for the Ru-doped Muonionalusta samples are summarized in Table S2. Experiment #1 (not doped with terrestrial Ru) should record the largest Ru anomalies and should reveal the true (but unknown) Ru isotopic composition of Muonionalusta. Importantly, a linear regression of all four experiments should run through the origin in a diagram of $\varepsilon^{i}Ru$ versus the relative fraction of meteoritical Ru in the experiment (Fig. S4, S5). The results of these experiments reveal good linear correlations for all $\varepsilon^{i}Ru$ values with the exception of $\varepsilon^{104}Ru$. However, the regression lines for $\varepsilon^{96}Ru$, $\varepsilon^{98}Ru$, $\varepsilon^{99}Ru$, and $\varepsilon^{104}Ru$ (all $^{102}Ru/^{100}Ru$-normalized) do not run through the origin and, rather, show positive intercepts (Fig. S4, S5), similar to the anomalies observed for the Ru-doped NIST SRM 129c measurements. Plotting the isotope data from the Muonionalusta experiments in the same manner as the Ru-doped NIST SRM 129c data in Fig. S3, reveals the same, yet somewhat more pronounced U-shape isotope pattern. Such patterns in internally normalized isotopic data are indicative of un-accounted-for fractionation effects introduced by non-exponential isotope fractionation in the analyzed samples (see Supplementary Information for more details).

Fischer-Gödde et al. (2015) reported a mean $\varepsilon^{96}Ru_{99/101}$ = 0.06±0.05 for several Ru-free reference materials doped with a terrestrial Ru standard. For their Ru-doped NIST SRM 129c measurements these authors found a mean $\varepsilon^{96}Ru_{99/101}$ = 0.10±0.10, similar to the mean $\varepsilon^{96}Ru_{99/101}$ = 0.16±0.05 obtained in the present study. Recalculating the data from Fischer-Gödde et al. (2015) for the $^{102}Ru/^{100}Ru$ normalization, the mean $\varepsilon^{96}Ru_{102/100}$ values of that study are 0.13±0.09 for all Ru-doped reference materials, and 0.22±0.17 for the Ru-doped NIST 129c analyses. For the latter we obtained an identical mean $\varepsilon^{96}Ru_{102/100}$ = 0.22±0.07 (all uncertainties are 95% conf.). Thus, the comparison of the Ru-doped NIST129c data suggests that unaccounted-for fractionation effects appear to be present in both studies with an approximatly similar magnitude.

*3.2 Iron meteorites*

The Ru isotope data for iron meteorites are reported in Table 1 and plotted in $\varepsilon^{100}Ru_{99/101}$ versus $\varepsilon^{96,98}Ru_{99/101}$ and $\varepsilon^{99}Ru_{102/100}$ versus $\varepsilon^{96,98}Ru_{102/100}$ diagrams in Fig. 1. All irons of this study exhibit negative $\varepsilon^{100}Ru_{99/101}$ and $\varepsilon^{102}Ru_{99/101}$ values and positive $\varepsilon^{96}Ru_{99/101}$ and $\varepsilon^{98}Ru_{99/101}$ values. Applying the $^{102}$Ru/$^{100}$Ru-normalization, all $\varepsilon^{i}Ru_{102/100}$ are positive and they decrease in the order $\varepsilon^{96}Ru_{102/100}$ > $\varepsilon^{98}Ru_{102/100}$ > $\varepsilon^{99}Ru_{102/100}$. Overall, these results are consistent with those of prior studies (Fischer-Gödde et al., 2015; Fischer-Gödde and Kleine, 2017; Bermingham et al., 2018; Worsham et al., 2019), but are more precise compared to prior MC-ICP-MS studies due to the larger amount of Ru analyzed for a single measurement. Significant intra-group variability for $\varepsilon^{100}Ru_{99/101}$ and $\varepsilon^{99}Ru_{102/100}$ for different samples of the same iron meteorite group is not observed, consistent with the expected absence of any CRE effects in the analyzed samples. However, some duplicate analyses (full sample replicates) do show offsets of up to ~0.6 $\varepsilon^{96}Ru_{99/101}$ and ~1 $\varepsilon^{96}Ru_{102/100}$, respectively (Fig. 1, Table 1). Regardless of these offsets, NC iron meteorites record well-correlated $\varepsilon^{100}Ru_{99/101}$ versus $\varepsilon^{96}Ru_{99/101}$ (or $\varepsilon^{99}Ru_{102/100}$ versus $\varepsilon^{96}Ru_{102/100}$) and $\varepsilon^{100}Ru_{99/101}$ versus $\varepsilon^{98}Ru_{99/101}$ (or $\varepsilon^{99}Ru_{102/100}$ versus $\varepsilon^{98}Ru_{102/100}$) variations, which are consistent with isotope variations expected from the heterogeneous distribution of *s*-process Ru (Fig. 1). By contrast, there are no resolved Ru isotope variations among the CC irons, which plot off the correlation lines defined by the NC irons.

Eight of the iron meteorites of this study (both NC and CC) were also analyzed by Fischer-Gödde et al. (2015), and there is very good agreement between these two studies for $\varepsilon^{99}Ru_{102/100}$ and $\varepsilon^{101}Ru_{102/100}$. On average the $\varepsilon^{96}Ru_{102/100}$ and $\varepsilon^{98}Ru_{102/100}$ values determined in the present study may be somewhat higher by 0.33±0.28 and 0.18±0.18, respectively, although this difference is not resolved (Fig. S6). Four of the iron meteorites of this study (Chihuahua City, Mount Dooling, Unter-Mässing, and Clark County) have also been analyzed by Worsham et al. (2019), and four other irons of this study (Toluca, Campo del Cielo, North Chile, and Chinga) were also analyzed by Bermingham and Walker (2017) and Bermingham et al. (2018) by TIMS. Comparison of these data to those of the present study reveal no systematic offsets, which at least in part may reflect the larger uncertainty of the mean $\varepsilon^{96}Ru_{99/101}$ and $\varepsilon^{98}Ru_{99/101}$ values obtained in these prior TIMS studies (Fig. S6).

### *3.3 Correction for non-exponential fractionation*

The effect of non-exponential fractionation may be quantified using the measured mass-dependent isotope fractionation of each processed sample (e.g., Tang and Dauphas, 2014). For the samples of this study, however, the mass-dependent isotope fractionation could not be measured because our analytical setup utilized an Aridus desolvator for sample introduction, which often causes disparate instrumental mass bias for sample and standard analyses. As such, this setup did not allow determining mass-dependent isotope variations by standard-sample bracketing. However, the data for the Ru-doped NIST SRM 129c and Muonionalusta measurements provide a means for correcting the iron meteorite data of this study for the effects of non-exponential fractionation. While the effect appears to be slightly larger for the doped Muonionalusta experiments than for NIST SRM 129c, this difference is not large and both samples show similar U-shaped Ru isotope patterns (Fig. S3). Consequently, to account for the non-exponential fractionation as well as currently possible, we used the mean $\varepsilon^{i}Ru$ values obtained for the Ru-doped NIST SRM 129c measurements and the intercepts of the Ru-doped Muonionalusta samples to correct the mean $\varepsilon^{i}Ru$ values of each iron meteorite group. The uncertainties of these $\varepsilon^{i}Ru$ values were quadratically added to the uncertainties of the measured group means (Table 1).

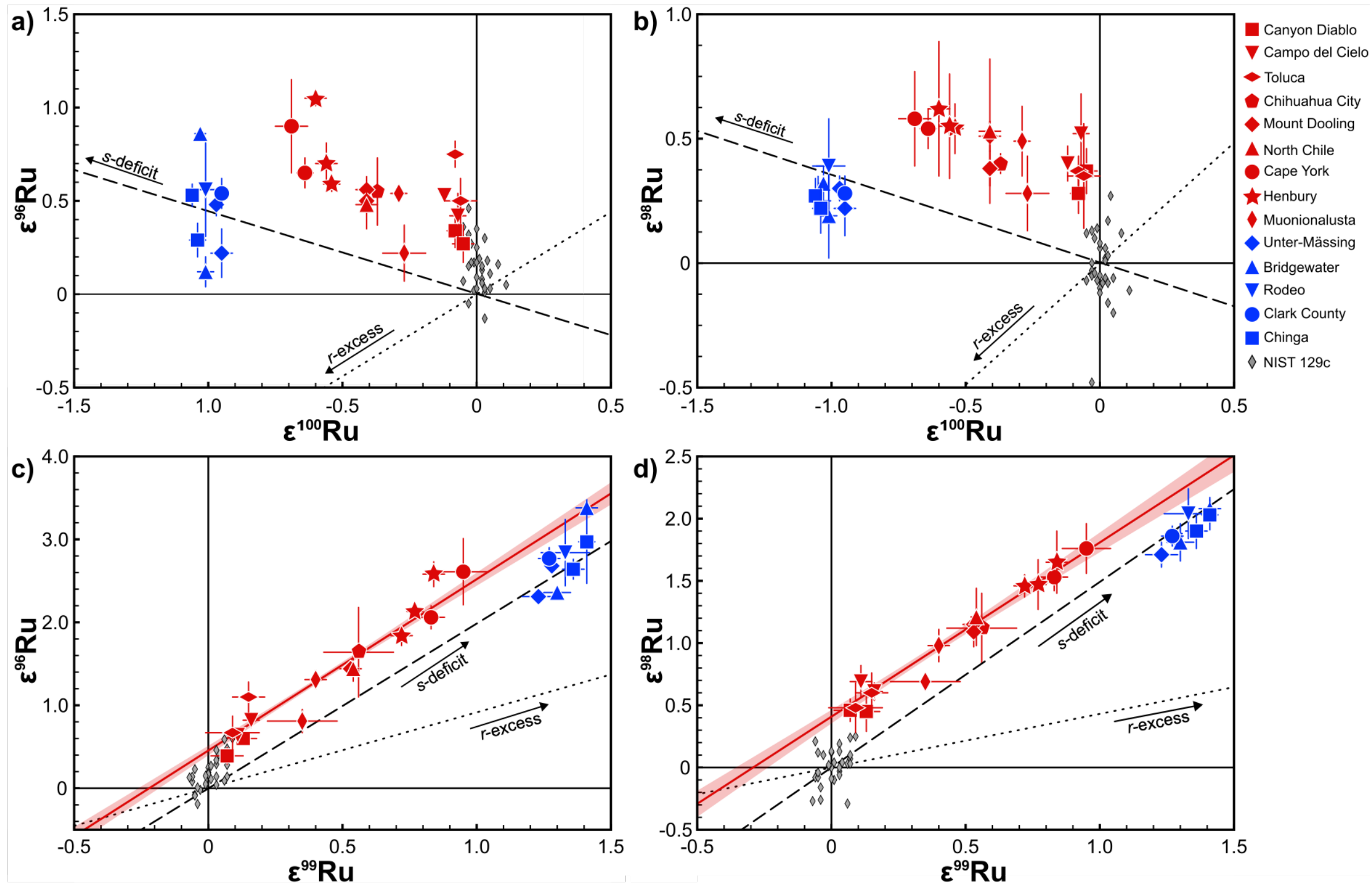

Fig. 1: Measured Ru isotope data of NC (red) and CC (blue) iron meteorites. The data are internally normalized to $^{99}Ru/^{101}Ru$ (a, b) or $^{102}Ru/^{100}Ru$ (c, d). The data reveal systematically distinct Ru isotope compositions of NC and CC iron meteorites. Also shown are the results of the Ru-doped NIST SRM 129c steel measurements (grey diamonds). The dashed and dotted lines indicate s- and r-process mixing lines, respectively, calculated from Ru isotope data for presolar SiC grains (Savina et al., 2004) and corresponding r-process residuals. The NC regression lines in (c) and (d) are calculated using IsoplotR (Vermeesch, 2018).

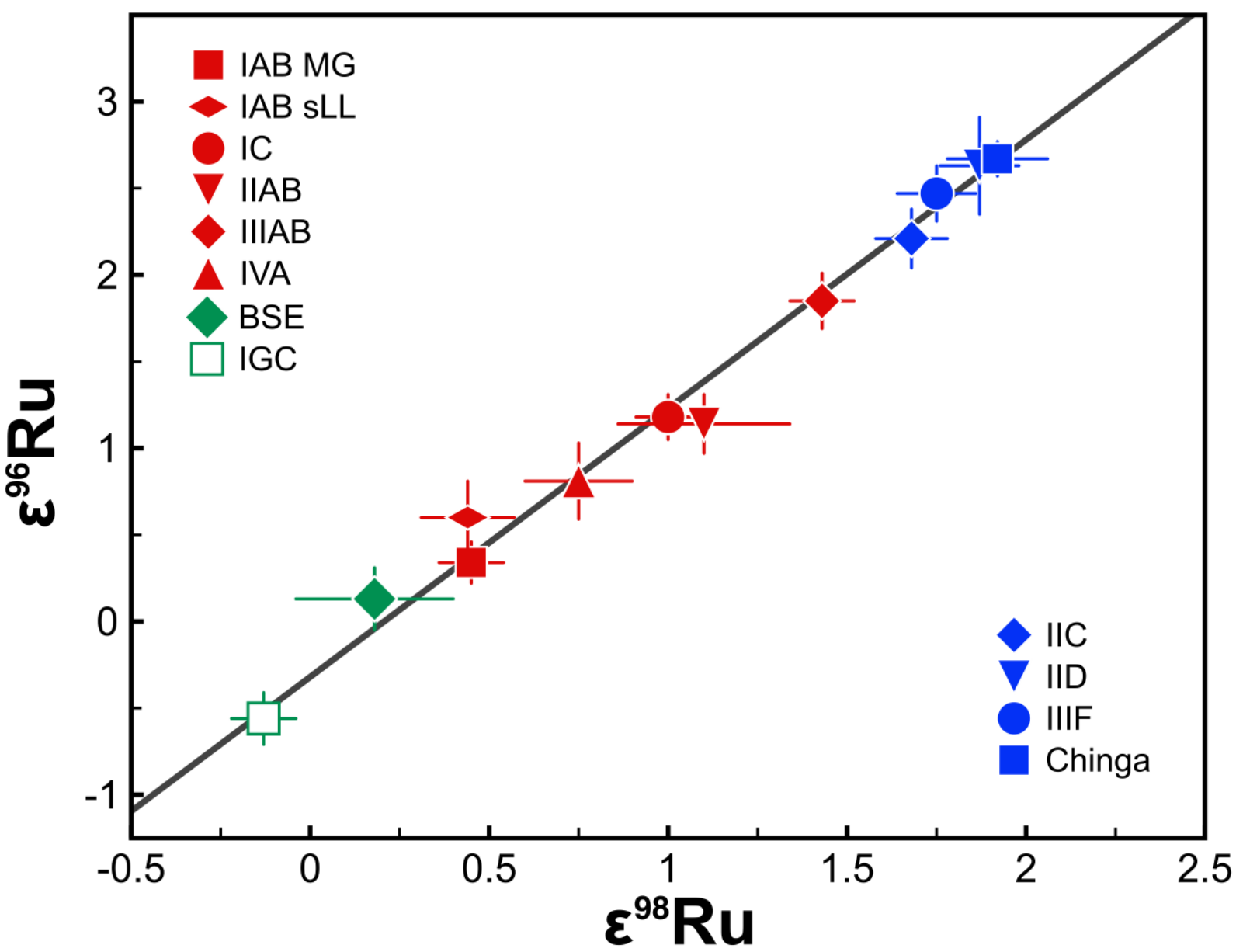

Fig. 2: Plot of $\varepsilon^{98}Ru_{102/100}$ versus $\varepsilon^{96}Ru_{102/100}$ after correction for effects of non-exponential fractionation as described in the text. Also shown are also data for the terrestrial Itsaq Gneis Complex (IGC) and the bulk silicate Earth (BSE). The black solid line is a regression through the iron meteorite data, revealing a well-defined correlation (MSWD = 0.8) of $\varepsilon^{98}$Ru and $\varepsilon^{96}$Ru.

The correction implicitly assumes that the unaccounted-for fractionation effect is approximately constant for all samples. Support for this assumption comes from the coherent and systematic Ru isotope variations we find among and between the NC and CC irons (Fig. 2). Importantly, the association of the irons to either NC or CC is based on their Mo isotope signatures and as such entirely independent of their Ru isotope composition measured in this study. Specifically, in a plot of $\varepsilon^{99}Ru_{102/100}$ versus $\varepsilon^{98}Ru_{102/100}$, the NC irons plot along a single correlation line, while all CC irons are systematically offset from this line. Moreover, in a plot of $\varepsilon^{96}Ru_{102/100}$ versus $\varepsilon^{98}Ru_{102/100}$, both NC and CC irons plot on a single well-defined correlation line with the CC irons displaying systematically larger anomalies compared to the NC irons. Such overall systematic behavior is not expected were the data compromised by variable analytical artefacts. Instead, such variable artefacts would result in unsystematic behavior with no clear distinction between NC and CC irons, and would result in scatter in the $\varepsilon^{96}Ru_{102/100}$ versus $\varepsilon^{98}Ru_{102/100}$ plot, which is not observed (Fig. 2).

We finally note that the correction for non-exponential fractionation is small for $\varepsilon^{98}Ru_{102/100}$ and in the $\varepsilon^{99}Ru_{102/100}$ versus $\varepsilon^{98}Ru_{102/100}$ plot results in shifts along the NC-line (Fig. 3), making this correction inconsequential for resolving the NC-CC dichotomy in this plot and for using it as a genetic tracer (see below).

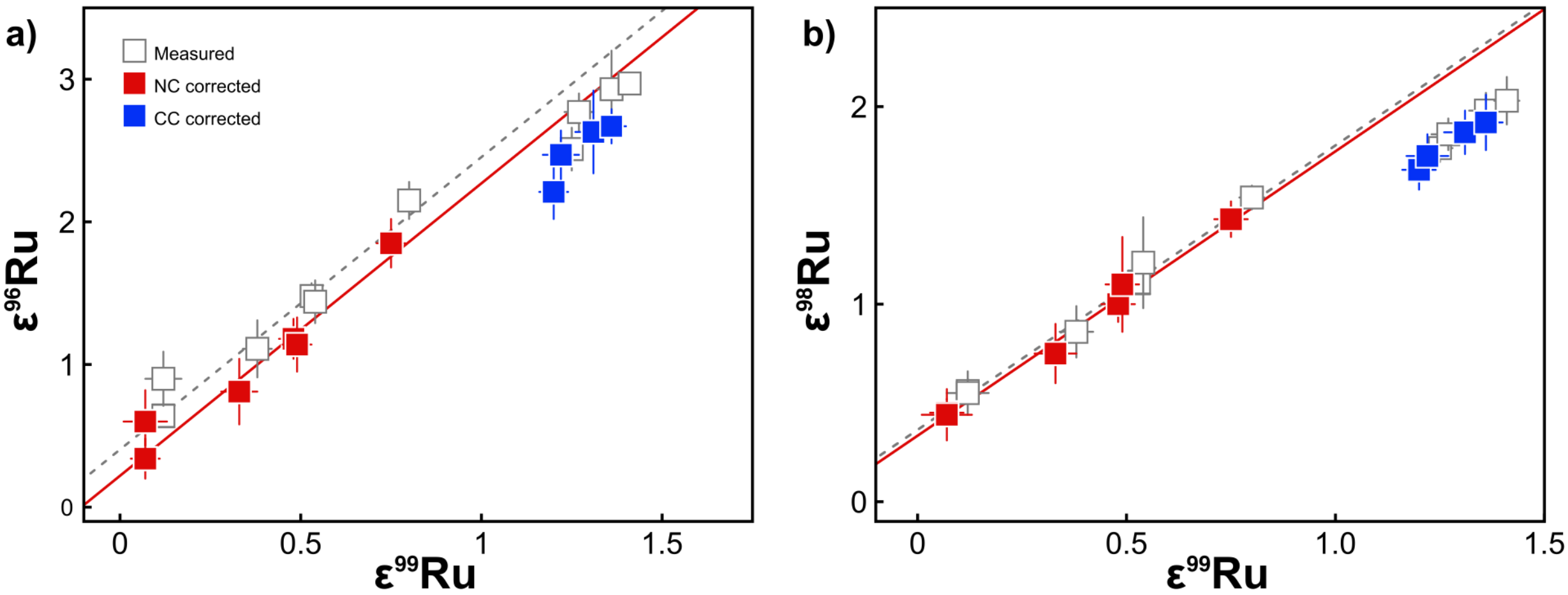

Fig. 3: Measured (grey symbols) versus corrected group mean values (colored symbols) for NC (red) and CC (blue) iron meteorites analyzed in this study. The NC regression lines were calculated using IsoplotR (Vermeesch, 2018); error envelopes are not shown for more clarity. The data are shown in (a) $\varepsilon^{99}Ru_{102/100}$–$\varepsilon^{96}Ru_{102/100}$ and (b) $\varepsilon^{99}Ru_{102/100}$–$\varepsilon^{98}Ru_{102/100}$ space. Because the fractionation effect is larger for $\varepsilon^{96}Ru$ than for $\varepsilon^{98}Ru$, the change of the position of the NC line is more prominent in $\varepsilon^{99}Ru_{102/100}$–$\varepsilon^{96}Ru_{102/100}$ space.

## 4. Discussion

### *4.1 Ru isotope dichotomy recorded in iron meteorites*

A key observation from the data of this study is that they reveal systematically distinct Ru isotope signatures for NC- and CC-type iron meteorites. This is evident in plots of either $\varepsilon^{100}Ru_{99/101}$ versus $\varepsilon^{96}Ru_{99/101}$ or $\varepsilon^{98}Ru_{99/101}$, or as $\varepsilon^{99}Ru_{102/100}$ versus $\varepsilon^{96}Ru_{102/100}$ or $\varepsilon^{98}Ru_{102/100}$ (Fig. 1, 4). In all these plots, the NC irons plot along a line, while the CC irons cluster off this line. As noted above, this offset cannot be explained by the unaccounted-for fractionation effect and, therefore, reflects a systematic difference in the Ru isotopic make-up of the NC and CC irons.

In the $\varepsilon^{100}Ru_{99/101}$ versus $\varepsilon^{96}Ru_{99/101}$ or $\varepsilon^{98}Ru_{99/101}$ diagrams, the heterogeneous distribution of *s*- and *r*-process Ru isotopes is expected to result in isotopic variations following two lines that are almost perpendicular to each other (Fig. 1). This situation is similar to the $\varepsilon^{94}Mo$-$\varepsilon^{95}Mo$ diagram and, theoretically, ideally suited for identifying a NC-CC dichotomy, because for Mo the CC reservoir is characterized by an approximately constant *r*-process excess over the NC reservoir (e.g., Budde et al., 2016). However, the CC irons have fairly homogeneous Ru isotope compositions, and so instead of defining a trend along an *s*-process mixing line, they plot in a distinct cluster (Bermingham et al., 2018; Worsham et al., 2019).

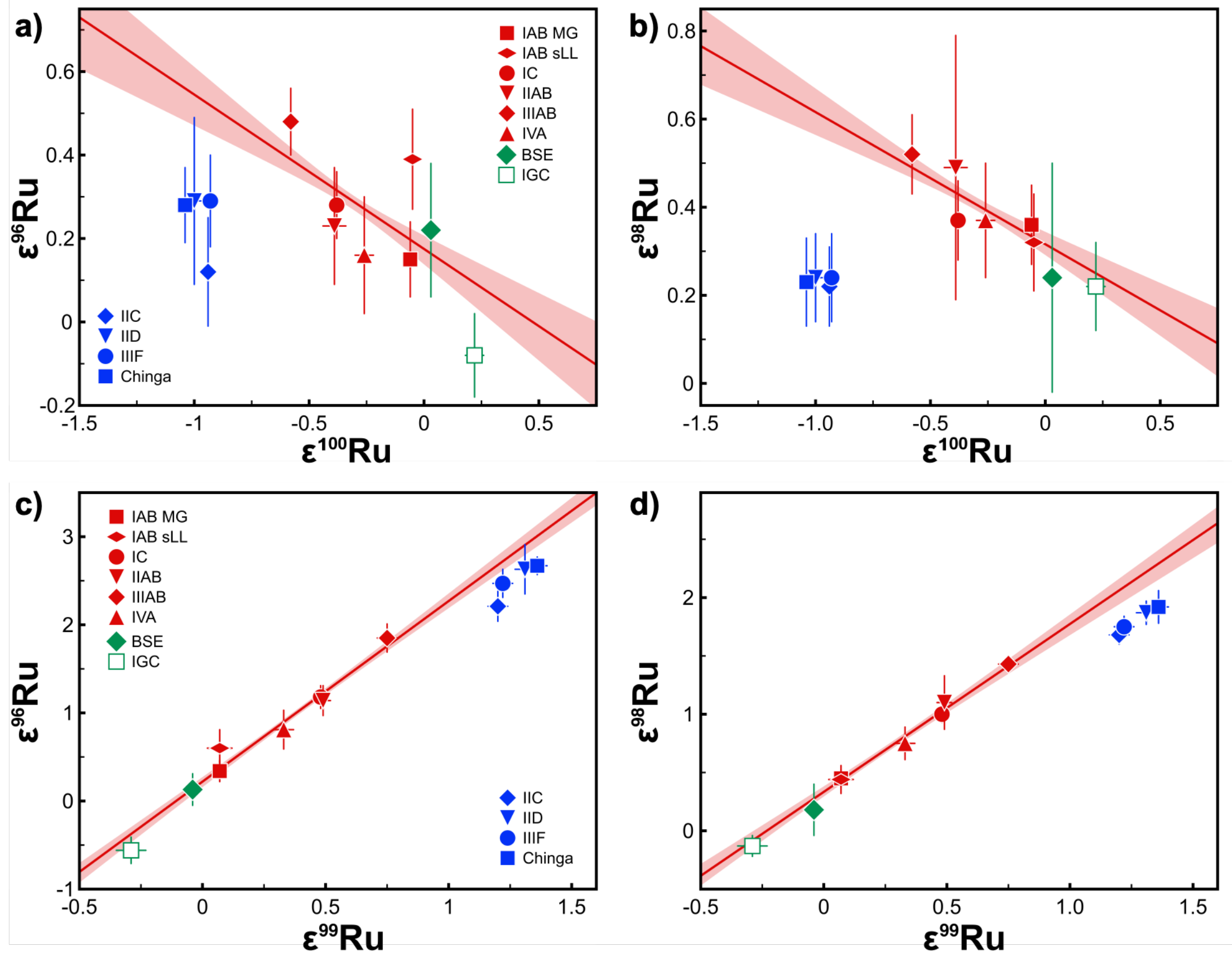

Fig. 4: Iron meteorite group mean values corrected for the unaccounted-for fractionation effect in the $^{99}Ru/^{101}Ru$- (a, b) and $^{102}Ru/^{100}Ru$-normalizations (c, d). The NC data plot along an *s*-process mixing line, and the CC iron meteorites are offset from this line. The NC regressions line (red) was calculated from the iron meteorite group mean values determined in this study using IsoplotR (Vermeesch, 2018). The BSE data are compiled from Fischer-Gödde et al. (2015), Fischer-Gödde and Kleine (2017), Bermingham and Walker (2017), and Fischer-Gödde et al., (2020) (see Table S4). The Itsaq gneiss complex data are from Fischer-Gödde et al. (2020). Uncertainties of the literature data represent the 95% confidence intervals of the mean for $N \geq 4$, and for $N < 4$ the 2 s.d. using the data of the original studies that have been re-cast into the $^{102}Ru/^{100}Ru$ normalization.

Moreover, for the $\varepsilon^{96}Ru_{99/101}$ and $\varepsilon^{98}Ru_{99/101}$ values there is no resolved difference between the NC and CC irons and any scattering of the data resulting from unaccounted-for fractionation is largely parallel to the y-axis of the diagram, so that the NC-CC isotopic difference is largely based on variations in $^{100}Ru$. This makes resolving the dichotomy using Ru three-isotope plots of $^{99}Ru/^{101}Ru$-normalized data less obvious (Fig. 1).

This situation is different for the alternative plots of $\varepsilon^{99}Ru_{102/100}$ versus $\varepsilon^{96}Ru_{102/100}$ or $^{98}Ru_{102/100}$ (Fig. 1, 4). For this normalization the anomalies are generally larger and the CC irons have more positive $\varepsilon^{96}Ru_{102/100}$ and $\varepsilon^{98}Ru_{102/100}$ values than the NC irons. Moreover, the NC irons show more extended isotopic variations along an *s*-process mixing line, and by this define the NC correlation line more precisely. The CC irons on the other hand plot off the NC-line by virtue of having more elevated $\varepsilon^{96}Ru_{102/100}$, $\varepsilon^{98}Ru_{102/100}$, and $\varepsilon^{99}Ru_{102/100}$ values, consistent with an approximately constant *r*-process Ru excess in CC over NC materials, as also observed for Mo isotopes.

In summary, a NC-CC dichotomy for Ru is apparent in three-isotope plots for both the (previously used) $^{99}Ru/^{101}Ru$ and the (newly introduced) $^{102}Ru/^{100}Ru$ normalizations. The latter is preferred here because it results in a more clearly defined NC-line and in resolved NC-CC isotopic differences for all three isotope ratios involved (i.e., $\varepsilon^{99}Ru_{102/100}$, $\varepsilon^{98}Ru_{102/100}$, and $\varepsilon^{96}Ru_{102/100}$). Moreover, the correction for non-exponential fractionation is smaller compared to the overall isotopic variations, and in particular in the $\varepsilon^{99}Ru_{102/100}$ versus $\varepsilon^{98}Ru_{102/100}$ plot results in (only small) shifts along the NC-line (Fig. 2), making this corrections inconsequential for resolving the NC-CC dichotomy in this plot.

### *4.2 Implications for the late veneer*

One of the key applications of the NC-CC dichotomy for Ru relates to determining the genetic heritage of the late veneer. Since the BSE's Ru predominantly derives from the late veneer, the position of the BSE's Ru isotopic composition with respect to the NC-line and the CC cluster can provide useful constraints on the contribution of NC and CC materials to late accretion. For the BSE, we use previously reported data from Bermingham and Walker (2017), which include samples from nine different oceanic and continental mantle domains, and combine them with data for terrestrital chromitites from Shetland and Bushveld (Fischer-Gödde et al., 2017; Fischer-Gödde et al., 2020). The Ru isotope data in these studies are reported using the $^{99}Ru/^{101}Ru$-normalization and for the comparsion to the data of this study they were recast into the $^{102}Ru/^{100}Ru$-normalization (see Table S4).There are additional Ru isotope data for terrestrial samples, but these are for samples with anomalous Ru isotopic compositions (Fischer-Gödde et al., 2020; Messling et al., 2025) and, as such, are not representative of the BSE. Our estimate of the BSE's Ru isotope composition is, therefore, based on the largest dataset of terrestrial samples currently available.

The analytical artefacts identified in this study introduce some uncertainty when combining $^{96}Ru$ and $^{98}Ru$ data from different sources, because these effects may vary among different studies and because unlike for the data of this study, no correction can be applied to the data from prior studies. Importantly, however, as noted above, these analytical effects result in correlated $\varepsilon^{99}Ru_{102/100}$ and $\varepsilon^{98}Ru_{102/100}$ variations along the NC-line and as such do not impact the distinction between NC and CC signatures in this plot (Fig. 2). Thus, although analytical artefacts may be present in the data for the terrestrial samples, these data can still be used to distinguish between and NC and CC origin of the terrestrial Ru. Moreover, for the $^{99}Ru/^{101}Ru$-normalization, the corrections for $\varepsilon^{98}Ru_{99/101}$ and $\varepsilon^{100}Ru_{99/101}$ are generally small and within uncertainty of the data. By contrast, for $\varepsilon^{96}Ru$ (both normalizations), the corrections are more significant and may result in deviations from the NC-line, which makes the comparison of data from different studies more uncertain in this case.

In the $\varepsilon^{99}Ru_{102/100}$ versus $\varepsilon^{98}Ru_{102/100}$ diagram (Fig. 4), the BSE ($\varepsilon^{98}Ru_{102/100} = 0.18 \pm 0.11$ and $\varepsilon^{99}Ru_{102/100} = 0.04 \pm 0.04$) plots on the NC-line and furthest away from the CC irons, which are characterized by the most elevated $\varepsilon^{98}Ru_{102/100}$ and $\varepsilon^{99}Ru_{102/100}$ values compared to the BSE. The same observation can be made from the $\varepsilon^{99}Ru_{102/100}$ versus $\varepsilon^{96}Ru_{102/100}$ diagram ($\varepsilon^{96}Ru_{102/100} = 0.13 \pm 0.18$) (Fig. 4), albeit with more scatter. For the $^{99}Ru/^{101}Ru$-normalization, the NC-line is far less well defined because the overall spread in $\varepsilon^{96}Ru_{99/101}$ and $\varepsilon^{98}Ru_{99/101}$ values is much smaller (Fig. 4). Nevertheless, as for the $^{102}Ru/^{100}Ru$-normalized data, the BSE plots at one end of the NC trend and furthest away from the CC irons. These observations suggest that the BSE's Ru and, hence, the late veneer is predominantly NC in origin and supports similar conclusions that were based on $\varepsilon^{100}Ru_{99/101}$ data alone (Fischer-Gödde and Kleine, 2017; Bermingham an Walker, 2017, Bermingham et al., 2018).

Several carbonaceous chondrites are known to have smaller $\varepsilon^{100}Ru_{99/101}$ anomalies than the CC irons (Fischer-Gödde et al., 2017), and so these samples also have lower $\varepsilon^{99}Ru_{102/100}$, $\varepsilon^{98}Ru_{102/100}$, and $\varepsilon^{96}Ru_{102/100}$, which may be closer to the BSE composition. Unfortunately, comparing the BSE's Ru isotope composition to those of carbonaceous chondrites is compromised by considerable uncertainty, because the $^{96}Ru$ and $^{98}Ru$ data for these samples are far less precise than those of the iron meteorites and in addition display substantial intra-group Ru isotope variability (Fischer-Gödde and Kleine, 2017). Nevertheless, despite these uncertainties, in the $\varepsilon^{99}Ru_{102/100}$ versus $\varepsilon^{96}Ru_{102/100}$ or $\varepsilon^{98}Ru_{102/100}$ plots the carbonaceous chondrites appear to be offset from the NC line in the same manner as the CC irons and none of them overlap with the composition of the BSE (Fig. S7).

Based on positive $\varepsilon^{100}Ru_{99/101}$ values for samples from the Itsaq Gneiss Complex (IGC) from Isua, Fischer-Gödde et al. (2020) argued for a significant contribution of CI chondrite-like material to the late veneer. These authors interpreted these rocks to partly have preserved the Ru isotope composition of a pre-late veneer mantle, because Ru might not have been fully extracted by core formation from the pre-late veneer mantle (Rubie et al., 2016). Since CI chondrites are characterized by negative $\varepsilon^{100}Ru_{99/101}$ values (i.e., positive $\varepsilon^{99}Ru_{102/100}$), the addition of such material would compensate for the positive $\varepsilon^{100}Ru_{99/101}$ (i.e., negative $\varepsilon^{99}Ru_{102/100}$) signatures observed at Isua (Fig. 4), resulting in a modern bulk BSE $\varepsilon^{100}Ru_{99/101}$ (and $\varepsilon^{99}Ru_{102/100}$) value of around zero. As shown in Fig. 4, the IGC composition, like the BSE, plots on the extension of the NC-line, indicating an NC origin of this material. As the BSE also plots on the NC-line, these data are therefore entirely consistent with a purely NC origin of the late veneer consisting of bodies having variable isotopic compositions along the NC-line.

However, given the current uncertainty on especially the $^{96}Ru$ and $^{98}Ru$ isotopic composition of the BSE, and the scatter in the overall $^{96}Ru$ data, a mixture between a mantle having an IGC-like Ru isotopic composition and CC material cannot be excluded. For the $\varepsilon^{99}Ru$-$\varepsilon^{96}Ru$ systematics (both normalizations), the contribution of CC material to the BSE's Ru appears less likely, but owing to the larger uncertainty on $\varepsilon^{96}Ru$ can also not be excluded (Fig. 4). Clearly, new high-precision $^{96}Ru$ and $^{98}Ru$ data for terrestrial samples are needed to better define the position of the BSE on the Ru three-isotope plots and, ultimately, to more quantitatively determine the potential contribution of CC material to the late veneer.

We nevertheless note that a significant contribution of CC material to the late veneer is neither suggested by the data, nor does it seem likely. Instead, the BSE plots on the NC-line defined in this study and between the IGC and NC meteorites, precisely as expected for a late veneer consisting of a heterogeneous mix of NC bodies. This interpretation is also consistent with the NC-like Ru-Mo isotopic data for lunar impactites, indicating that late-accreted bodies hitting the Moon are predominantly NC and isotopically most similar to enstatite chondrites (Worsham and Kleine, 2021). Lunar samples do not appear to record the full complement of the late veneer, as the lunar mantle likely started retaining highly siderophile elements only after ~4.35 Ga (Morbidelli et al., 2018), ~150 Ma after its formation at ~4.5 Ga (e.g., Mezger et al., 2021; Nimmo et al., 2024; Schneider et al., 2025). All data combined are thus, consistent with a scenario in which the late veneer initially consisted of NC bodies having *s*-process-enriched isotope signatures as recorded in the IGC and over time evolved to more *s*-process-depleted signatures as recorded in NC meteorites.

## 5. Conclusions

New high-precision Ru isotope data for iron meteorites reveal ubiquitous analytical artefacts predominantly on $^{96}$Ru (and to a lesser extent on $^{98}$Ru), which likely reflect non-exponential isotope fractionation introduced during sample preparation. Despite these effects, the new data demonstrate that in three-isotope plots that involve the *p*-process Ru nuclides $^{96}$Ru or $^{98}$Ru, NC and CC meteorites have systematically distinct compositions. This NC-CC dichotomy for Ru is most readily resolved in plots of $\varepsilon^{99}Ru_{102/100}$ versus $\varepsilon^{96}Ru_{102/100}$ or $\varepsilon^{98}Ru_{102/100}$, in which the NC irons plot along a line reflecting variations in the abundance of *s*-process Ru. The CC irons are offset from this line, reflecting an approximately constant *r*-process Ru excess in the CC over the NC reservoir. These systematics are nearly identical to those for Mo isotopes.

The presence of an NC-CC dichotomy in Ru three-isotope plots offers a new tool for investigating the genetic heritage of the late veneer. The new iron meteorite data combined with previously reported Ru isotope data for terrestrial samples indicate that terrestrial materials plot along the NC-line and show no detectable contribution of CC materials to the BSE's Ru. This suggests a predominantly NC origin of the late veneer, which may have evolved over time from a more *s*-process-enriched composition as recorded in the Itsaq Gneiss Complex to a more *s*-process-depleted composition as recorded in NC meteorites. However, in light of the analyical artefacts identified in this study, new high-precision $^{96}$Ru and $^{98}$Ru isotope data for terrestrial samples and carbonaceous chondrites are needed to more rigorously assess any contribution of primitive carbonaceous chondrite-like materials to the late veneer.

**Data availability**

All data used in this manuscript are provided in the main text and in the Supplementary material.

**Acknowledgements**

We gratefully acknowledge constructive comments from Richard Walker and an anonymous reviewer, the efficient editorial handlying by Paolo Sossi, and helpful discussion with Jan Render and Jan Hellmann. Funded by the Deutsche Forschungsgemeinschaft (DFG, German Research Foundation) – Project-ID 263649064 – TRR 170. A portion of this work was performed under the auspices of the U.S. Department of Energy by Lawrence Livermore National Laboratory under Contract DE-AC52-07NA27344.

**Appendix A. Supplementary Material**

The Supplementary Material provides all supplementary figures and tables referenced in the main text. Figure S2 shows the results of Zr and Mo doping experiments on the accuracy of the Zr and Mo interference corrections on $\varepsilon^{i}$Ru. Figure S3-S5 presents the results for the Ru-doped NIST_129c and Muonionalusta Experiments. Figure S6 shows a comparison of Ru isotope data for iron meteorites obtained in this study with literature data from Fischer-Gödde et al. (2015), Worsham et al. (2019), and Bermingam et al. (2018) for the same samples. Figure S7 shows an extesion of Fig. 4 including data for chondrites. The Supplementary Tables S1 and S2 contain Ru isotope data of the NIST_129c and Muonionalusta experiments, and Table S3 shows Ru data for single measurements of the Bridgewater replicate sample solution applying cone setups #1 and #2. Table S4 contains the Ru isotope data of terrestrial sample used for calculation of the bulk silicate Earth (BSE) values.

# References

Becker, H., Horan, M. F., Walker, R. J., Gao, S., Lorand, J. P., Rudnick, R. L., 2006. Highly siderophile element composition of the Earth's primitive upper mantle: constraints from new data on peridotite massifs and xenoliths. Geochim. Cosmochim. Acta, 70, 4528-4550.

Bermingham, K.R., Walker, R.J., Worsham, E.A., 2016. Refinement of high precision Ru isotope analysis using negative thermal ionization mass spectrometry. International journal of mass spectrometry, 403, 15-26.

Bermingham, K.R., Walker, R.J., 2017. The ruthenium isotopic composition of the oceanic mantle. Earth Planet. Sci. Lett. 474, 466-473.

Bermingham, K.R., Worsham, E.A., Walker, R.J., 2018. New insights into Mo and Ru isotope variation in the nebula and terrestrial planet accretionary genetics. Earth Planet. Sci. Lett. 487, 221-229.

Bermingham, K.R., Füri, E., Lodders, K., Marty, B., 2020. The NC-CC isotope dichotomy: Implications for the chemical and isotopic evolution of the early Solar System. Space science reviews 216, 1-29.

Bermingham, K.R., Tornabene, H.A., Walker, R.J., Godfrey, L.V., Meyer, B.S., Piccoli, P., Mojzsis, S.J., 2025. The non-carbonaceous nature of Earth's late-stage accretion. Geochim. Cosmochim. Acta 392, 38-51.

Birck, J.L., Barman, M.R., Capmas, F., 1997. Re-Os isotopic measurements at the femtomole level in natural samples. Geostandards newsletter, 21(1), 19-27.

Brenan, J. M., McDonough, W. F., 2009. Core formation and metal–silicate fractionation of osmium and iridium from gold. Nature Geoscience, 2(11), 798-801.

Budde, G., Burkhardt, C., Brennecka, G.A., Fischer-Gödde, M., Kruijer, T.S., Kleine, T., 2016. Molybdenum isotopic evidence for the origin of chondrules and a distinct genetic heritage of carbonaceous and noncarbonaceous meteorites. Earth Planet. Sci. Lett. 454, 293–303.

Budde, G., Burkhardt, C., Kleine, T., 2019. Molybdenum isotopic evidence for the late accretion of outer Solar System material to Earth. Nature Astronomy 3(8), 736-741.

Burkhardt, C., Spitzer, F., Morbidelli, A., Budde, G., Render, J. H., Kruijer, T. S., Kleine, T., 2021. Terrestrial planet formation from lost inner solar system material. Sci. Adv. 7 (52).

Calvo, L., Siebert, J., Huang, D., Blanchard, I., Kubik, E., Bonino, V., Schreiber, A., Avice, G., Labidi, J., 2026. Accretion of volatile elements on Earth without the need of a late veneer. Science advances, 12(9), eady8018.

Chen, J.H., Papanastassiou, D.A., Wasserburg, G.J., 2010. Ruthenium endemic isotope effects in chondrites and differentiated meteorites. Geochim. Cosmochim. Acta 74(13), 3851-3862.

Chou, C. L., 1978. Fractionation of siderophile elements in the earth's upper mantle. In: *Lunar and Planetary Science Conference*, *vol*. 9. pp. 219–230.

Corrigan, C.M., Nagashima, K., Hilton, C., McCoy, T. J., Ash, R.D., Tornabene, H. A., Walker, R.J., McDounough, W.F., Rumble, D., 2022. Nickel-rich, volatile depleted iron meteorites: Relationships and formation processes. Geochim. Cosmochim. Acta, 333, 1-21.

Dauphas, N., Davis, A.M., Marty, B., Reisberg, L., 2004. The cosmic molybdenum–ruthenium isotope correlation. Earth Planet. Sci. Lett. 226(3-4), 465-475.

Dauphas, N., Schauble, E.A., 2016. Mass fractionation laws, mass-independent effects, and isotopic anomalies. Annual Review of Earth and Planetary Sciences, 44(1), 709-783.

Dauphas, N., 2017. The isotopic nature of the Earth's accreting material through time. Nature 541(7638), 521-524.

Ek, M., Hunt, A.C., Lugaro, M., Schönbächler, M., 2020. The origin of s-process isotope heterogeneity in the solar protoplanetary disk. Nature Astronomy, 4(3), 273-281.

Fischer-Gödde, M., Burkhardt, C., Kruijer T.S., Kleine, T., 2015. Ru isotope heterogeneity in the solar protoplanetary disk. Geochim. Cosmochimi. Acta 168, 151-171.

Fischer-Gödde, M., Kleine, T., 2017. Ruthenium isotopic evidence for an inner Solar System origin of the late veneer. Nature 541(7638), 525-527.

Fischer-Gödde, M., Elfers, B.M., Münker, C., Szilas, K., Maier, W.D., Messling, N., Morishita, T., Van Kranendonk, M., Smithies, H., 2020. Ruthenium isotope vestige of Earth's pre-late-veneer mantle preserved in Archaean rocks. Nature 579(7798), 240-244.

Fischer-Gödde, M., Tusch, J., Goderis, S., Bragagni, A., Mohr-Westheide, T., Messling, N., Elfers B.-M., Schmitz B., Reimold W.U., Maier W., Claeys P., Koeberl C., Tissot F.L.H., Bizzarro M., Münker C., 2024. Ruthenium isotopes show the Chicxulub impactor was a carbonaceous-type asteroid. Science, 385(6710), 752-756.

Hopp, T., Fischer-Gödde, M., Kleine, T., 2018. Ruthenium isotope fractionation in protoplanetary cores. Geochimica et Cosmochimica Acta, 223, 75-89.

Kleine, T., Budde, G., Burkhardt, C., Kruijer, T.S., Worsham, E.A., Morbidelli, A., Nimmo, F., 2020. The non-carbonaceous–carbonaceous meteorite dichotomy. Space Science Reviews 216, 1-27.

Kruijer, T.S., Fischer-Gödde, M., Kleine, T., Sprung, P., Leya, I., Wieler, R., 2013. Neutron capture on Pt isotopes in iron meteorites and the Hf–W chronology of core formation in planetesimals. Earth Planet. Sci. Lett. 361, 162-172.

Kruijer, T.S., Burkhardt, C., Budde, G., Kleine, T., 2017. Age of Jupiter inferred from the distinct genetics and formation times of meteorites. Proceedings of the National Academy of Sciences 114, 6712–6716.

Mezger, K., Maltese, A., and Vollstaedt, H., 2021. Accretion and differentiation of early planetary bodies as recorded in the composition of the silicate Earth. Icarus, 365, 114497.

Messling, N., Willbold, M., Kallas, L., Elliott, T., Fitton, J. G., Müller, T., and Geist, D., 2025. Ru and W isotope systematics in ocean island basalts reveals core leakage. Nature, 642, 376-380.

Morbidelli, A., Nesvorny, D., Laurenz, V., Marchi, S., Rubie, D. C., Elkins-Tanton, L., Wieczorek, M., and Jacobson, S., 2018. The timeline of the lunar bombardment: Revisited. Icarus, 305, 262-276.

Nimmo, F., Kleine, T., and Morbidelli, A., 2024. Tidally driven remelting around 4.35 billion years ago indicates the Moon is old. Nature, 636, 598-602.

Pape, J., Zhang, B., Spitzer, F., Rubin, A.E., Kleine, T., 2024 Isotopic constraints on genetic relationships among group IIIF iron meteorites, Fitzwater Pass, and the Zinder pallasite. Meteorit. Planet. Sci. 59, 778–788.

Poole, G.M., Rehkämper, M., Coles, B.J., Goldberg, T., Smith, C.L., 2017. Nucleosynthetic molybdenum isotope anomalies in iron meteorites–new evidence for thermal processing of solar nebula material. Earth Planet. Sci. Lett. 473, 215–226.

Prantzos, N., Abia, C., Cristallo, S., Limongi, M., Chieffi, A., 2020. Chemical evolution with rotating massive star yields II. A new assessment of the solar s-and r-process components. Monthly Notices of the Royal Astronomical Society, 491(2), 1832-1850.

Qin, L., Carlson, R.W., 2016. Nucleosynthetic isotope anomalies and their cosmochemical significance. Geochemical Journal, 50(1), 43-65.

Righter, K., Humayun, M., Danielson, L., 2008. Partitioning of palladium at high pressures and temperatures during core formation. Nature Geoscience, 1(5), 321-323.

Rubie, D. C., Laurenz, V., Jacobson, S. A., Morbidelli, A., Palme, H., Vogel, A. K., & Frost, D. J. (2016). Highly siderophile elements were stripped from Earth's mantle by iron sulfide segregation. *Science*, *353*(6304), 1141-1144.

Savina, M.R., Davis, A.M., Tripa, C.E., Pellin, M.J., Gallino, R., Lewis, R.S., Amari, S., 2004. Extinct technetium in silicon carbide stardust grains: implications for stellar nucleosynthesis. Science 303(5658), 649-652.

Schneider, J. M., and Kleine, T., 2025. The age and early evolution of the Moon revealed by the Rb-Sr systematics of lunar ferroan anorthosites. Earth and Planetary Science Letters, 669, 119592.

Sossi, P. A., and Bower, D. J., 2026. Homogeneous accretion of the Earth in the inner Solar System. Nature Astronomy, 1-8.

Spitzer, F., Burkhardt, C., Budde, G., Kruijer, T.S., Morbidelli, A., Kleine, T., 2020. Isotopic evolution of the inner solar system inferred from molybdenum isotopes in meteorites. The Astrophysical Journal Letters, 898(1), L2.

Spitzer, F., Burkhardt, C., Pape, J., Kleine, T., 2022. Collisional mixing between inner and outer solar system planetesimals inferred from the Nedagolla iron meteorite. Meteorit. Planet. Sci. 57, 261-276.

Tornabene, H. A., Hilton, C. D., Bermingham, K. R., Ash, R. D., and Walker, R. J., 2020. Genetics, age and crystallization history of group IIC iron meteorites. Geochim. Cosmochim. acta, 288, 36-50.

Trinquier, A., Birck, J.L., Allegre, C.J., 2007. Widespread $^{54}$Cr heterogeneity in the inner solar system. The Astrophysical Journal, 655(2), 1179.

Vermeesch, P., 2018. IsoplotR: A free and open toolbox for geochronology. Geoscience Frontiers 9(5), 1479-1493.

Walker, R.J., 2009. Highly siderophile elements in the Earth, Moon and Mars: update and implications for planetary accretion and differentiation. Geochemistry 69(2), 101-125.

Warren, P.H., 2011. Stable-isotopic anomalies and the accretionary assemblage of the Earth and Mars: A subordinate role for carbonaceous chondrites. Earth Planet. Sci. Lett. 311, 93-100.

Worsham, E.A., Bermingham, K.R., Walker, R.J., 2017. Characterizing cosmochemical materials with genetic affinities to the Earth: Genetic and chronological diversity within the IAB iron meteorite complex. Earth Planet. Sci. Lett. 467, 157–166.

Worsham, E.A., Burkhardt, C., Budde, G., Fischer-Gödde, M., Kruijer, T.S., Kleine, T., 2019. Distinct evolution of the carbonaceous and non-carbonaceous reservoirs: Insights from Ru, Mo, and W isotopes. Earth Planet. Sci. Lett. 521, 103–112.

Worsham, E.A., Kleine, T., 2021. Late accretionary history of Earth and Moon preserved in lunar impactites. Science advances 7(44).

Yokoyama, T., Nagai, Y., Fukai, R., Hirata, T., 2019. Origin and evolution of distinct molybdenum isotopic variabilities within carbonaceous and noncarbonaceous reservoirs. The Astrophysical Journal, 883(1), 62.

Icarus

## Supplementary Material for

# The NC-CC dichotomy in ruthenium isotopes: Implications for the origin of the late veneer

Jonas Pape[1], Emily Worsham[1,2], and Thorsten Kleine[1,3]

[1]Institut für Planetologie, University of Münster, Wilhelm-Klemm-Str. 10, 48149 Münster, Germany

[2]Nuclear and Chemical Sciences Division, Lawrence Livermore National Laboratory, Livermore, CA, USA,

[3]Max Planck Institute for Solar System Research, Justus-von-Liebig-Weg 3,37077 Göttingen, Germany

## 1. Ru purification

All samples were digested in 20 – 30 ml in-house, double-distilled 6 M HCl for several hours at 120 °C. After complete digestion, the samples were cooled down and the same volume of in-house distilled concentrated $HNO_3$ was added to produce reverse aqua regia in order to first oxidize the sample prior to subsequent reduction in HCl, which was important to minimize Ru loss in the first cation column described below. The solutions were placed on a hot plate overnight at 130 °C and subsequently evaporated to near dryness at <80 °C in order to suppress evaporative loss of Ru. The samples were then re-dissolved and evaporated to near dryness at temperatures <120 °C twice in 6 M HCl and once in 1 M HCl. Finally, the samples were re-dissolved in 1M HCl, diluted with Milli-Q water to 0.2 M HCl and loaded onto 0.7 x 30 cm (12 ml) BioRad Econo-Columns® filled with 10 ml pre-cleaned and conditioned BioRad AG 50W-X8 resin (100-200 mesh), where the Ru and other HSEs are separated from major and minor matrix elements like Fe, Ni and Cr. Ruthenium was eluted with 9 ml double-distilled 0.2 M HCl. A maximum of 0.3 g of iron meteorite was loaded onto single columns and larger samples were split over several columns and recombined before the subsequent micro-distillation. Aliquots of the sample solutions were analyzed before and after the first cation column by quadrupole ICP-MS in order to determine Ru yields and concentrations of matrix elements. Depending on the efficiency of the first cation column, some samples were cleaned by a second round of cation chemistry. This second cation column step is a miniature version of the first cation column, has similarly high Ru yields as the former, and was only necessary for the few samples for which the Fe capacity of the first cation column was exceeded due to sample weights close to 0.3 g. For each sample, the Ru from multiple columns was re-combined and dried down at <80 °C, re-dissolved in three drops of reverse aqua regia in order to oxidize and destroy organics, evaporated to near dryness, diluted with 50 µl of Milli-Q water and transferred to the lid of conical 5ml Savillex beakers, and dried at <80 °C for micro-distillation (Birck et al., 1997; Fischer-Gödde et al., 2015). During the micro-distillation, the samples were covered with 50 µl di-chromate ($H_2SO_4$-$CrO_3$) solution (i.e., 0.2 g $CrO_3$ per 1 ml of 6 M $H_2SO_4$), and stepwise heated on a hotplate. First the samples were slowly heated to 60 °C, then after ca. 12 h the temperature was raised to 70 °C, and after another ca. 10 h, to 80 °C, where the samples were kept for about 24 h before decreasing the temperature first to 70 °C and afterwards slowly to room temperature. During this procedure, the Ru was evaporated into a droplet (ca. 18 µl) of conc. HBr in the top of the micro-distillation vial. After micro-distillation the HBr containing the purified Ru fraction was dried down and taken up in 0.28M $HNO_3$ for isotope analysis.

The Ru yields of single steps and of the whole purification process were determined based on Ru concentrations for the samples that either are reported in the literature or were determined using quadrupole ICP-MS on digestion aliquots. The yields after the cation columns were typically >95%, while the yields of the micro-distillation were variable and ranged between ~43 and >90 %, which is typical for micro-distillation of Ru (Fischer-Gödde et al., 2015; Bermingham et al., 2016; Fischer-Gödde et al., 2024). As in previous studies, other than keeping the different parameters of the micro-distillation as constant as possible (i.e., amounts and concentrations of acids used, the heating and cooling path during the experiment) we were not able to identify the primary factor(s) controlling the varying yields of the micro-distillation. One parameter that might have significant impact upon the Ru yield of the micro-distillation could be the amount and type of sample matrix, as it was observed that the yield of micro-distillation of meteorite samples was generally higher compared to Ru-doped NIST steels, and significantly higher than for the pure Alfa Aesar Specpure™ analytical Ru standard.

## 2. Fractionation effects in internally normalized Ru isotope data

The U-shaped isotope patterns of the Ru-doped NIST SRM 129c and Muonionalusta measurements are best accounted for by mass-dependent isotope fractionation effects, which are inadequately corrected-for by the internal mass bias correction because they are not solely following the exponential fractionation law. Prior studies have shown that such incomplete correction of any analytical or natural mass-dependent fractionation can led to spurious mass-independent isotope anomalies, which, when misinterpreted as nucleosynthetic isotope anomalies, can lead to wrong conclusions (Dauphas and Schauble, 2016; Budde et al. 2023; Bermingham et al., 2025; Fitoussi et al., 2025). Using the formalism of Tang and Dauphas (2014) applied to $^{102}$Ru/$^{100}$Ru-normalized Ru isotope data we are able to reproduce the Ru isotope pattern observed for the Ru-doped NIST SRM 129c and Muonionalusta measurements for modest isotope fractionation of between 0.2 and 0.4 ‰ and for $n = -1$ (i.e., equilibrium instead of kinetic isotope fractionation as assumed in the exponential law) (Fig. S2). Unfortunately though, we were unable to precisely measure the mass-dependent isotope fractionation for any of the samples, because the observed fractionations are entirely dominated by relatively large instrumental fractionation introduced through the use of the Aridus as the sample introduction system.

The unaccounted-for fractionation effects may rather result from processes during sample preparation. Although we cannot exclude contribution from sample digestion and column chemistry, for Ru the unaccounted-for fractionation effects are suspected to be introduced during the micro-distillation, because here the Ru undergoes complex chemical reactions and it is this stage of the Ru purification process that can be least controlled. While the Ru yields of the micro-distillation were highly variable for the NIST SRM 129c experiments, we did not observe a systematic correlation between the yield and the magnitude of fractionation. Additionally, it was not possible to reduce or eliminate the fractionation effects by adjusting single parameters of the micro-distillation, like the heating and cooling curves or the amounts of acids used during the distillation procedure. Similarly, the Muonionalusta experiments show that the magnitude of the fractionation effect does not correlate with the yield achieved during the sample purification and micro-distillation. This is because Muonionalusta experiments #1 to #4 have increasing micro-distillation yields between 43 and 93%, respectively. If the fractionation effect were directly related to the Ru yields, this would result in curves rather than linear correlations when plotting $\varepsilon^{i}$Ru *versus* the relative fraction of meteoritical Ru in the experiment, which is not the case (Fig. S3, S4). Thus, despite the variable Ru yields, all four samples display approximately similar offsets (correlating with the fraction of meteoritical Ru in the experiment), suggesting that, if micro-distillation is the source of the un-accounted for fractionation, yields even slightly below 100% already introduce an analytical artefact and that this artefact does not systematically increase further with a worsening yield.

It is possible to omit the micro-distillation from the Ru-purification procedure to minimize the risk of introducing fractionation issues connected to this step by applying additional rounds of ion chromatography (Hopp et al., 2020; Worsham and Kleine, 2021). However, this is not straightforward when studying the Ru isotopic dichotomy, because in contrast to the former studies, here it is necessary to also acquire high-precision data for the lower abundance isotopes $^{96}$Ru or $^{98}$Ru. This requires both processing larger amounts of sample and producing extremely clean final sample solutions to minimize interference corrections. To our knowledge, no chromatographic procedure for Ru purification is available to date that would meet this criteria.

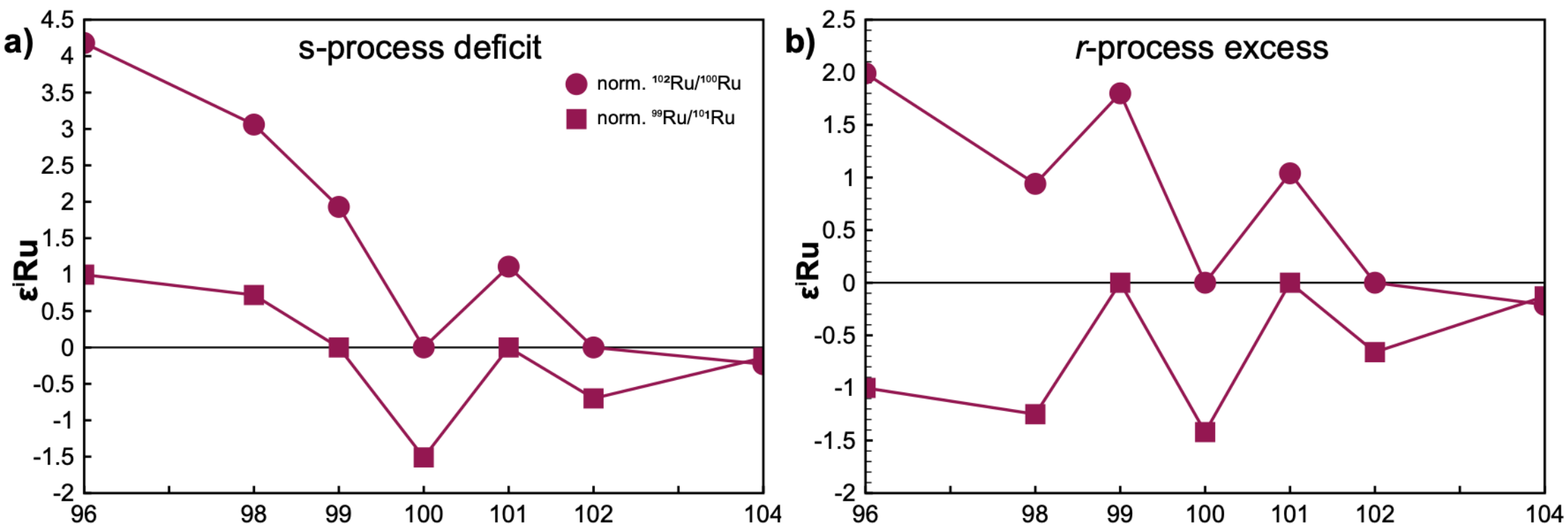

Fig. S1: Comparison of the Ru isotope pattern for a) *s*-process deficit and b) *r*-process excess relative to terrestrial Ru when using either $^{99}$Ru/$^{101}$Ru or $^{102}$Ru/$^{100}$Ru for internal normalization of the Ru isotope data and scaled to $\varepsilon^{96}$Ru = 1 (a) or $\varepsilon^{96}$Ru = -1 (b) for the $^{99}$Ru/$^{101}$Ru normalization. The nucelaosynthetic origins of each isotope are discussed in the main text.

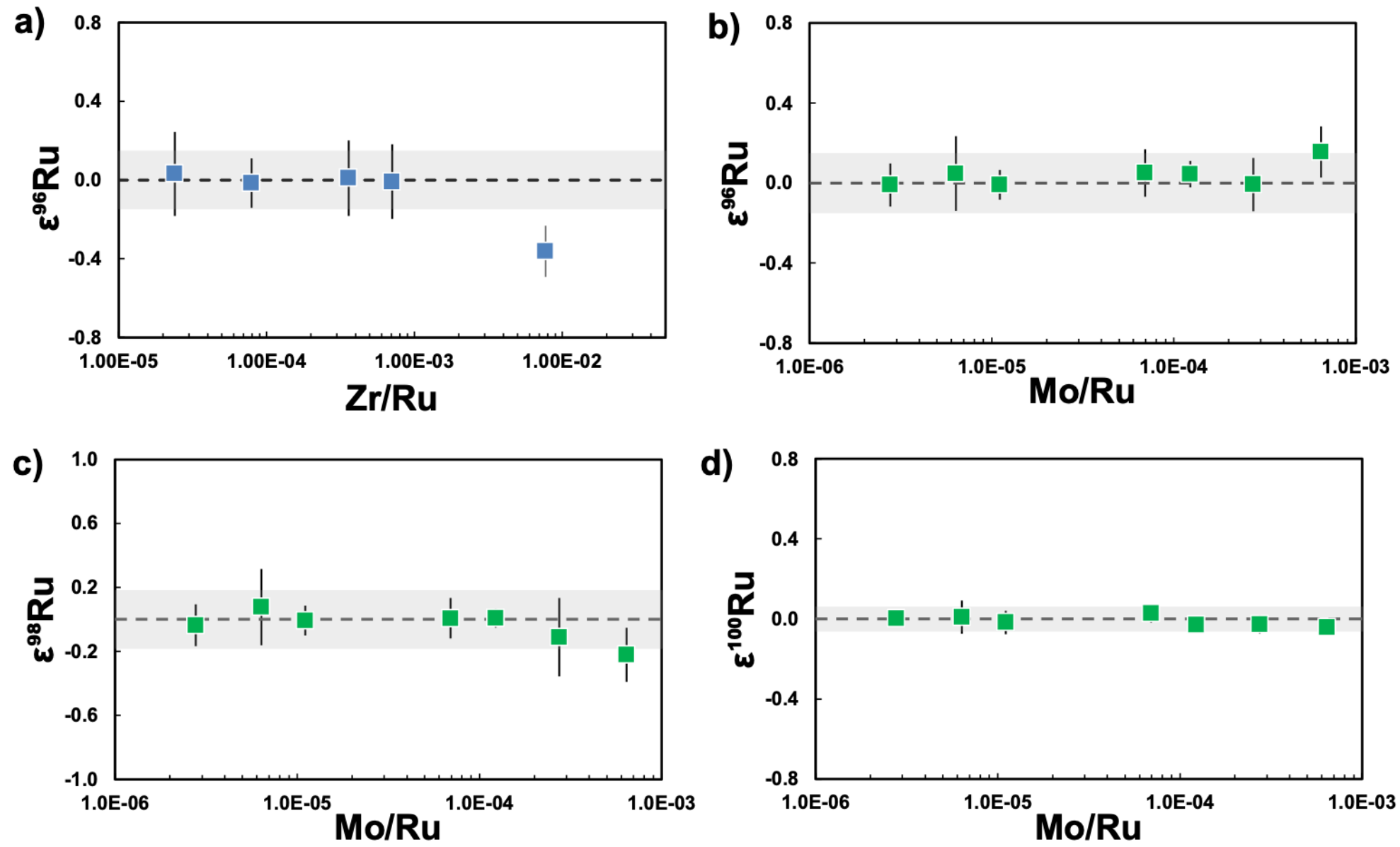

Fig. S2: Effects of Zr (a) and Mo (b - d) interference on the measured $\varepsilon^{96}$Ru, $\varepsilon^{98}$Ru, and $\varepsilon^{100}$Ru were tested by doping a Ru standard solution with various amounts of Zr or Mo, showing that the interference corrections are accurate for Zr/Ru ratios of up to $8 \times 10^{-4}$ and Mo/Ru ratios of up to $2 \times 10^{-4}$. The reader is referred to the main text for a detailed discussion.

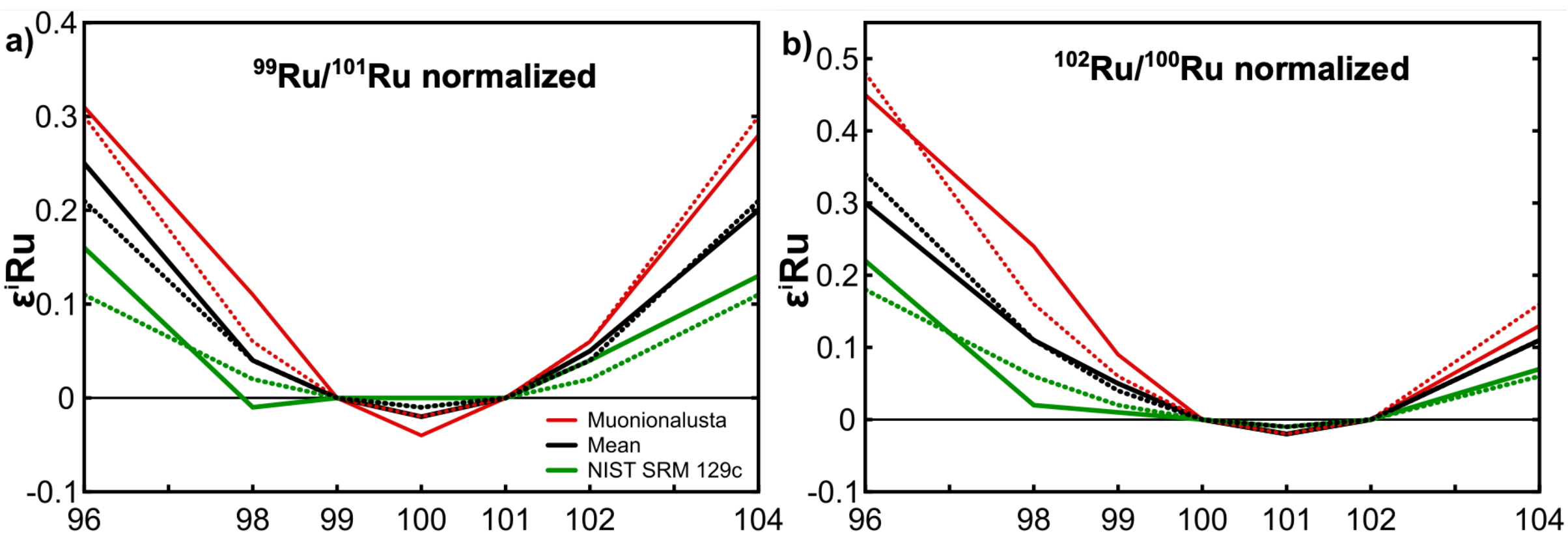

Fig. S3: Measured (solid lines) and modeled (dotted lines) Ru isotope fractionation for Ru-doped NIST SRM 129c steel (green), iron meteorite Muonionalusta (red) and the calculated mean (black). The Muonionalusta data correspond to the intercepts of the respective data shown in Figs. S4 and S5. The samples were processed through the whole Ru purification procedure and reveal U-shape patterns around the normalizing isotope ratios $^{99}$Ru/$^{101}$Ru (a) and $^{102}$Ru/$^{100}$Ru (b), indicating fractionation effects that are not resolved during internal mass bias correction using the exponential law. Error bars are omitted for clarity.

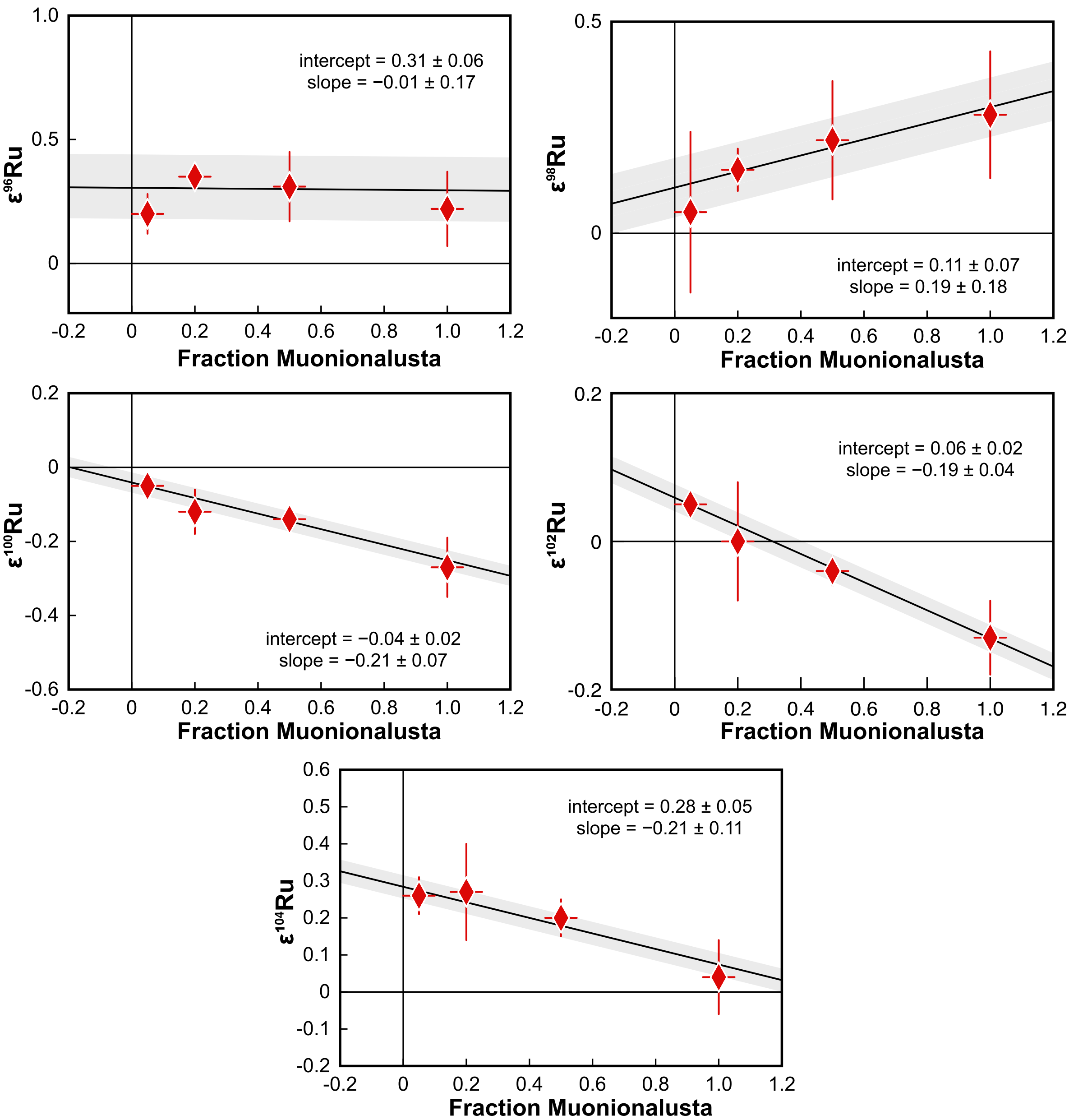

Fig. S4: Experiments of iron meteorite Muonionalusta that was doped with different amounts of terrestrial Ru standard solution prior to Ru separation and purification by ion chromatography and micro-distillation. The x-axis reflects the calculated fraction of (meteoritical) Ru in the experiment deriving from iron meteorite Muonionalusta over the Ru added through terrestrial standard solution prior to sample processing. The data were internally normalized to $^{99}Ru/^{101}Ru$ and reveal similar fractionation effects as were observed with the experiments using Ru-doped NIST SRM 129c steel (Fig. S3).

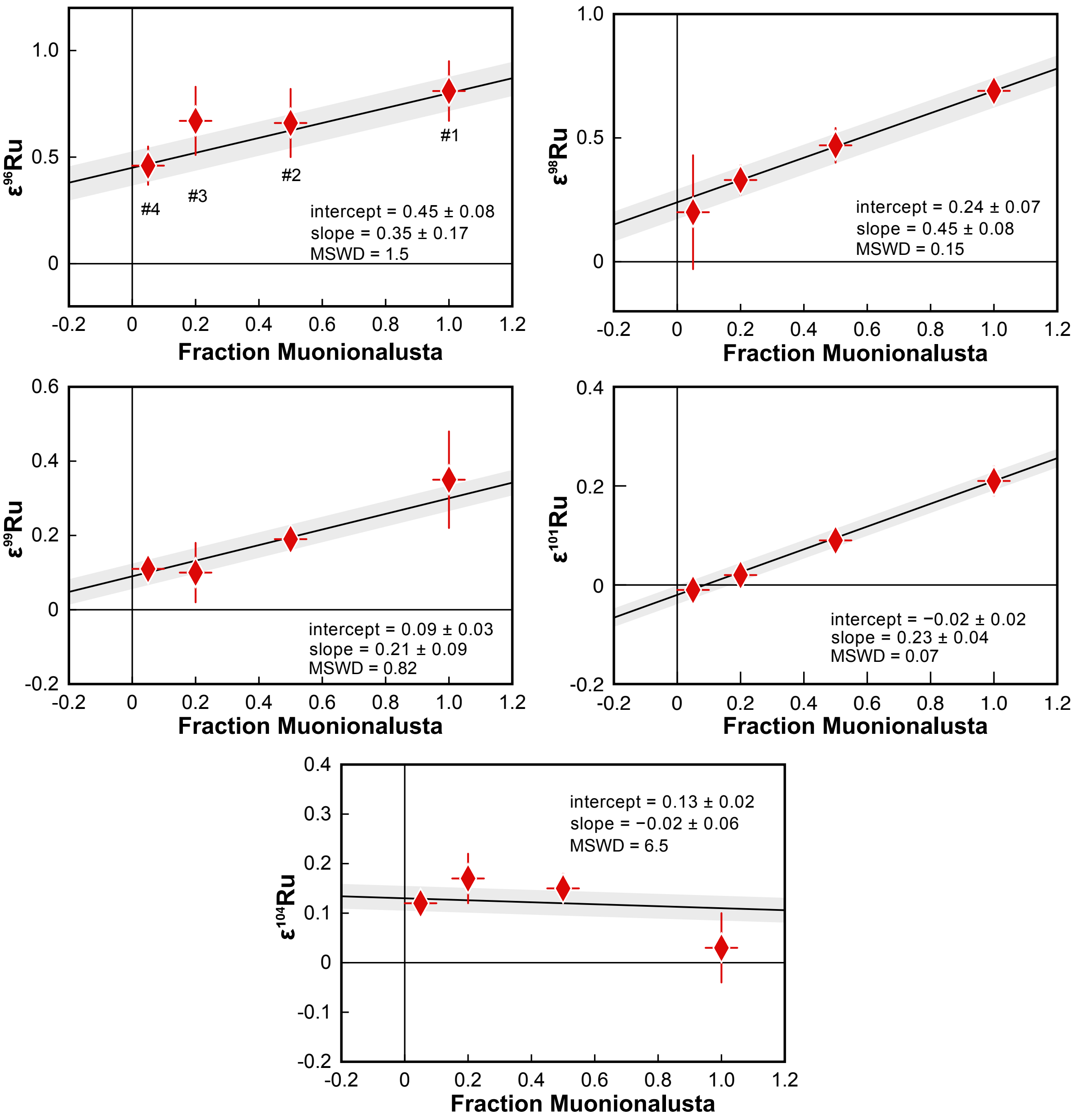

Fig. S5: Experiments #1 - #4 (denoted in the $\varepsilon^{96}$Ru diagram) of iron meteorite Muonionalusta that was doped with different amounts of the terrestrial Ru standard prior to Ru separation and purification by ion chromatography and micro-distillation. The x-axis reflects the calculated fraction of (meteoritical) Ru in the experiment deriving from the iron meteorite Muonionalusta. The data were internally normalized to $^{102}$Ru/$^{100}$Ru and reveal similar fractionation effects as were observed with the experiments using Ru-doped NIST SRM 129c steel (Fig. S3).

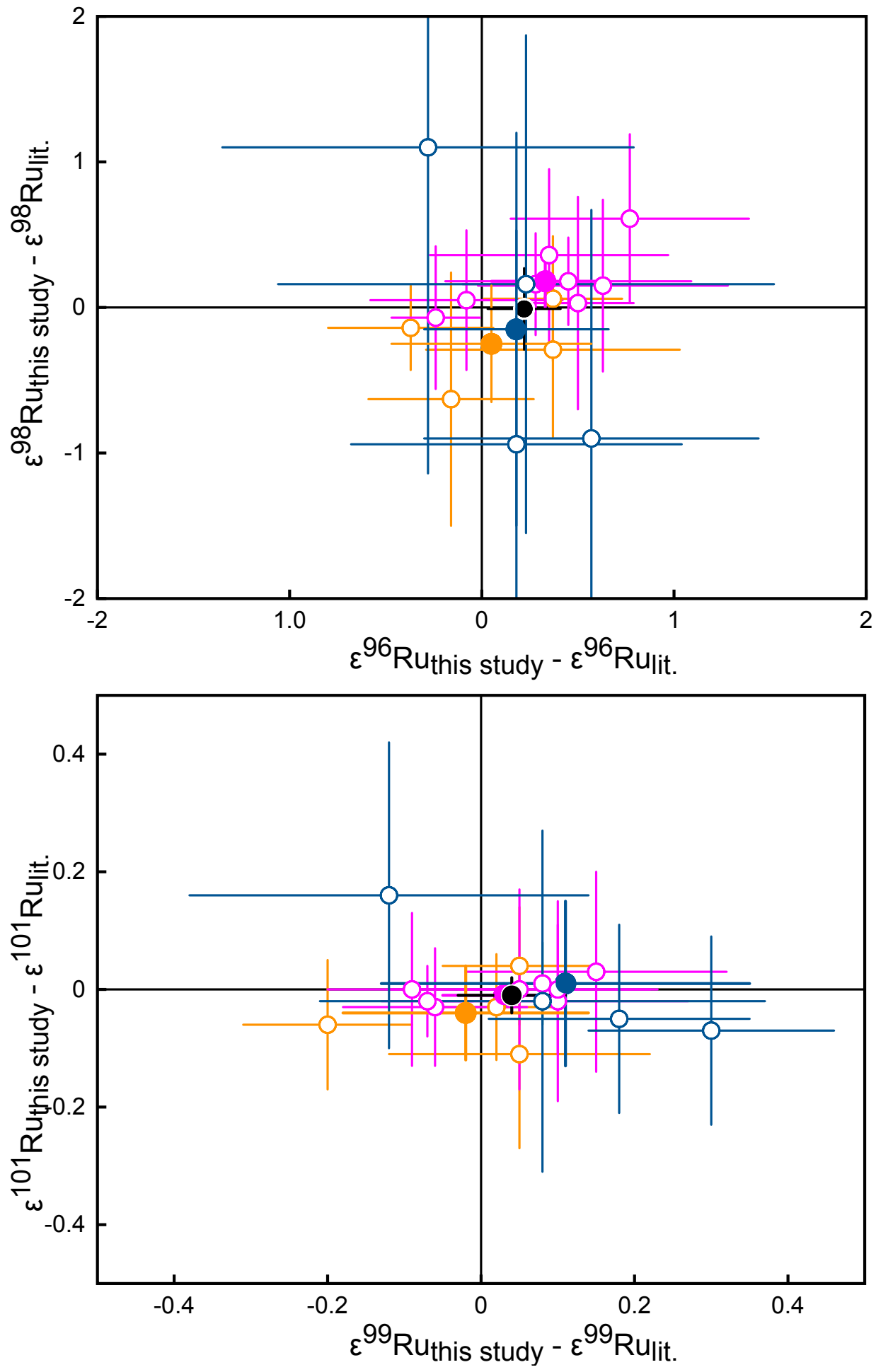

Fig. S6: The comparison of Ru isotope data for eight iron meteorites obtained in this study with literature data (Fischer-Gödde et al., 2015 (open pink symbols); Worsham et al, 2019 (open orange symbols); Bermingam et al., 2018 (open blue symbols)) for the same samples (average values for each data set are filled symbols, total average is black circle, all with 95% CI) show very good agreement between these studies for most of the $\varepsilon^{99}Ru_{102/100}$ and all $\varepsilon^{101}Ru_{102/100}$, but also reveals that while the $\varepsilon^{96}Ru_{102/100}$ and $\varepsilon^{98}Ru_{102/100}$ values generally agree within their respective uncertainties, both values tend to be slightly more positive in this study compared to the Fischer-Gödde et al. (2015) data, while the $\varepsilon^{98}Ru_{102/100}$ values tend to be more negative in this study compared to the data from Worsham et al. (2019) and Bermingam et al. (2018).

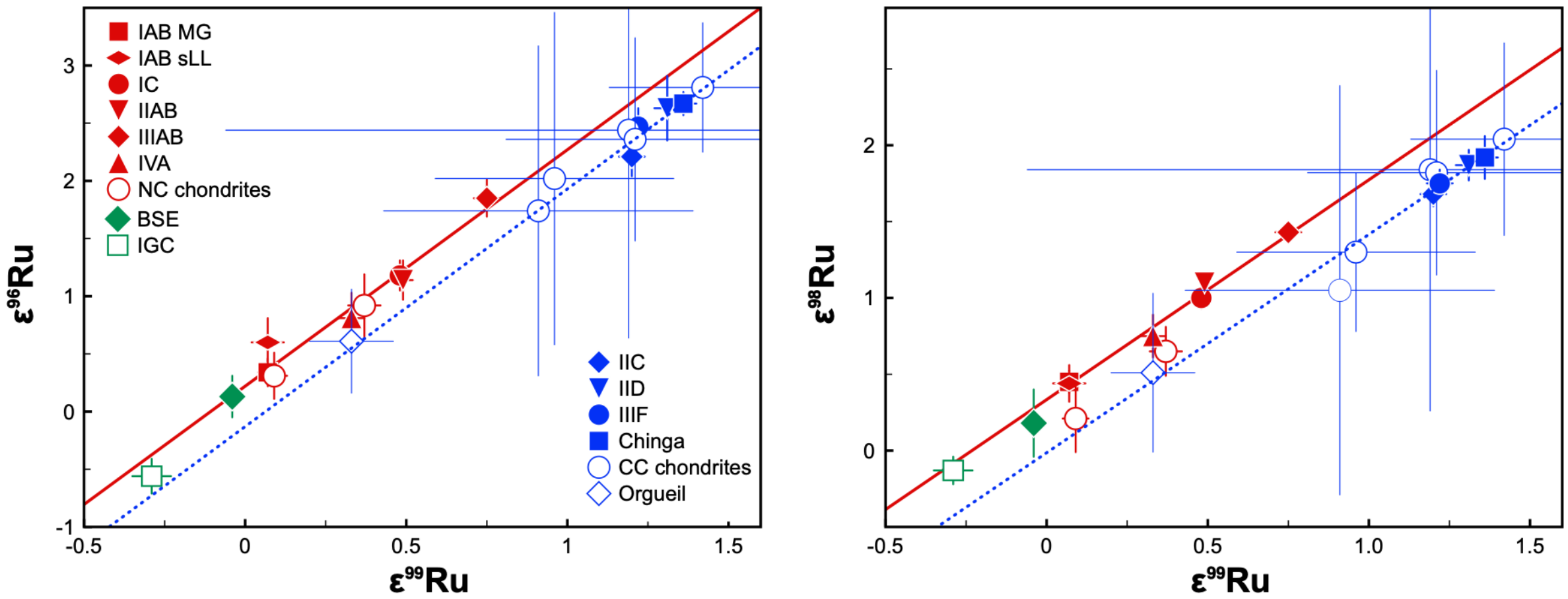

Fig. S7: $\varepsilon^{99}Ru_{102/100}$–$\varepsilon^{96}Ru_{102/100}$ (a) and $\varepsilon^{99}Ru_{102/100}$–$\varepsilon^{98}Ru_{102/100}$ (b) plots similar to Fig. 4, but now also including data for carbonaceous chondrites from Fischer-Gödde and Kleine (2017), recalculated to the $^{102}Ru/^{100}Ru$ normalisation. The NC regressions line (red) was calculated from the iron meteorite group mean values determined in this study using IsoplotR (Vermeesch, 2018). The CC trend line (blue dotted line) was calculated from the CC iron meteorite group mean values from this study and the CC chondrite data from Fischer-Gödde and Kleine (2017). Within their comparatively large uncertainties, the carbonaceous chondrites overlap with both the NC- and CC-lines.

TABLE S1. Ru isotope data for Ru-doped reference material NIST 129c steel.

| Sample | N | Normalized to $^{99}Ru/^{101}Ru$ | | | | | Normalized to $^{102}Ru/^{100}Ru$ | | | | |
|---|---|---|---|---|---|---|---|---|---|---|---|
| | | $\varepsilon^{96}$Ru (± 2σ) | $\varepsilon^{98}$Ru (± 2σ) | $\varepsilon^{100}$Ru (± 2σ) | $\varepsilon^{102}$Ru (± 2σ) | $\varepsilon^{104}$Ru (± 2σ) | $\varepsilon^{96}$Ru (± 2σ) | $\varepsilon^{98}$Ru (± 2σ) | $\varepsilon^{99}$Ru (± 2σ) | $\varepsilon^{101}$Ru (± 2σ) | $\varepsilon^{104}$Ru (± 2σ) |
| NIST 129c A1 [a)] | 3 | 0.01 ± 0.13 | -0.21 ± 0.47 | 0.03 ± 0.04 | -0.01 ± 0.06 | 0.09 ± 0.17 | 0.09 ± 0.10 | -0.08 ± 0.37 | 0.06 ± 0.05 | 0.04 ± 0.04 | -0.01 ± 0.10 |
| NIST 129c A2 [a)] | 3 | 0.11 ± 0.17 | 0.17 ± 0.21 | 0.03 ± 0.02 | 0.03 ± 0.02 | 0.07 ± 0.17 | 0.09 ± 0.21 | 0.14 ± 0.11 | -0.03 ± 0.06 | -0.03 ± 0.04 | 0.05 ± 0.10 |
| NIST 129c A3 [a)] | 3 | 0.11 ± 0.11 | -0.06 ± 0.33 | 0.08 ± 0.06 | 0.09 ± 0.16 | 0.03 ± 0.17 | 0.13 ± 0.23 | -0.07 ± 0.38 | -0.05 ± 0.02 | -0.09 ± 0.10 | 0.05 ± 0.10 |
| NIST 129c B1 [a)] | 3 | 0.09 ± 0.15 | 0.01 ± 0.23 | 0.02 ± 0.06 | 0.03 ± 0.01 | 0.06 ± 0.17 | 0.12 ± 0.36 | 0.01 ± 0.21 | -0.02 ± 0.05 | -0.03 ± 0.06 | 0.07 ± 0.10 |
| NIST 129c B2 [a)] | 3 | 0.17 ± 0.27 | -0.05 ± 0.28 | 0.00 ± 0.05 | 0.03 ± 0.04 | 0.15 ± 0.17 | 0.19 ± 0.18 | -0.04 ± 0.47 | 0.00 ± 0.11 | -0.02 ± 0.03 | 0.08 ± 0.30 |
| NIST 129c B3 [a)] | 3 | 0.12 ± 0.14 | -0.06 ± 0.14 | 0.01 ± 0.02 | 0.05 ± 0.09 | 0.12 ± 0.17 | 0.11 ± 0.12 | -0.06 ± 0.11 | 0.00 ± 0.02 | -0.03 ± 0.07 | -0.01 ± 0.10 |
| NIST 129c C1 [b)] | 2 | 0.20 ± 0.15 | 0.04 ± 0.07 | 0.01 ± 0.03 | 0.03 ± 0.00 | 0.13 ± 0.02 | 0.24 ± 0.27 | 0.07 ± 0.16 | 0.01 ± 0.06 | -0.02 ± 0.05 | 0.08 ± 0.01 |
| NIST 129c C2 [b)] | 2 | 0.31 ± 0.11 | -0.04 ± 0.01 | -0.01 ± 0.03 | 0.04 ± 0.08 | 0.24 ± 0.04 | 0.43 ± 0.07 | 0.02 ± 0.05 | 0.04 ± 0.01 | -0.01 ± 0.04 | 0.16 ± 0.10 |
| NIST 129c C3 [b)] | 2 | 0.37 ± 0.25 | -0.04 ± 0.02 | -0.03 ± 0.00 | 0.05 ± 0.03 | 0.21 ± 0.07 | 0.52 ± 0.19 | 0.03 ± 0.01 | 0.06 ± 0.01 | -0.02 ± 0.01 | 0.09 ± 0.04 |
| NIST 129c C4 [b)] | 2 | 0.16 ± 0.03 | 0.04 ± 0.10 | -0.02 ± 0.04 | 0.03 ± 0.01 | 0.17 ± 0.03 | 0.28 ± 0.02 | 0.07 ± 0.09 | 0.04 ± 0.09 | 0.01 ± 0.02 | 0.10 ± 0.11 |
| NIST 129c C5 [b)] | 2 | -0.05 ± 0.23 | -0.03 ± 0.17 | 0.03 ± 0.01 | 0.01 ± 0.01 | 0.00 ± 0.06 | -0.14 ± 0.13 | -0.10 ± 0.16 | -0.04 ± 0.01 | -0.03 ± 0.01 | 0.02 ± 0.13 |
| NIST 129c C6 [b)] | 2 | 0.34 ± 0.05 | 0.13 ± 0.03 | -0.04 ± 0.02 | 0.07 ± 0.00 | 0.26 ± 0.11 | 0.54 ± 0.17 | 0.24 ± 0.01 | 0.08 ± 0.03 | -0.02 ± 0.03 | 0.14 ± 0.01 |
| Mean | | 0.16 ± 0.25 | -0.01 ± 0.19 | 0.00 ± 0.06 | 0.04 ± 0.05 | 0.13 ± 0.17 | 0.22 ± 0.40 | 0.02 ± 0.20 | 0.01 ± 0.09 | -0.02 ± 0.06 | 0.07 ± 0.10 |
| 95% conf. | | 0.05 | 0.04 | 0.01 | 0.01 | 0.03 | 0.07 | 0.04 | 0.02 | 0.01 | 0.02 |

[a)] Aridus II, conventional Ni H-cones

[b)] Aridus 3, X-skimmer + H-sampler cones

TABLE S2. Measured Ru isotope data for iron meteorite Muonionalusta that was doped with terrestrial Ru standard solution (Muonionalusta #2-#4).

| Sample | N | Normalized to $^{99}Ru/^{101}Ru$ | | | | | Normalized to $^{102}Ru/^{100}Ru$ | | | | | |
|---|---|---|---|---|---|---|---|---|---|---|---|---|
| | | $\varepsilon^{96}Ru$ (± 2σ) | $\varepsilon^{98}Ru$ (± 2σ) | $\varepsilon^{100}Ru$ (± 2σ) | $\varepsilon^{102}Ru$ (± 2σ) | $\varepsilon^{104}Ru$ (± 2σ) | $\varepsilon^{96}Ru$ (± 2σ) | $\varepsilon^{98}Ru$ (± 2σ) | $\varepsilon^{99}Ru$ (± 2σ) | $\varepsilon^{101}Ru$ (± 2σ) | $\varepsilon^{104}Ru$ (± 2σ) | Fraction Muonionalu |
| ***Muonionalusta #1*** | 4 | 0.22 ± 0.15 | 0.28 ± 0.15 | -0.27 ± 0.08 | -0.13 ± 0.05 | 0.04 ± 0.10 | 0.81 ± 0.14 | 0.69 ± 0.02 | 0.35 ± 0.1 | 0.21 ± 0.03 | 0.03 ± 0.07 | 1 |
| ***Muonionalusta #2*** | 4 | 0.31 ± 0.14 | 0.22 ± 0.14 | -0.14 ± 0.03 | -0.04 ± 0.02 | 0.20 ± 0.05 | 0.66 ± 0.16 | 0.47 ± 0.07 | 0.19 ± 0.03 | 0.09 ± 0.02 | 0.15 ± 0.03 | 0.5 |
| ***Muonionalusta #3*** | 4 | 0.35 ± 0.05 | 0.15 ± 0.05 | -0.12 ± 0.06 | 0.00 ± 0.08 | 0.27 ± 0.13 | 0.67 ± 0.16 | 0.33 ± 0.06 | 0.10 ± 0.08 | 0.02 ± 0.02 | 0.17 ± 0.05 | 0.2 |
| ***Muonionalusta #4*** | 4 | 0.20 ± 0.08 | 0.05 ± 0.19 | -0.05 ± 0.02 | 0.05 ± 0.01 | 0.26 ± 0.05 | 0.46 ± 0.09 | 0.20 ± 0.23 | 0.11 ± 0.03 | -0.01 ± 0.02 | 0.12 ± 0.02 | 0.05 |

Uncertainties represent the 95% confidence intervals of the mean (i.e., (s.d. ×t0.95, N-1)/√N).

TABLE S3. Ru isotope data for single measurements of the Bridgewater replicate sample solution using cone setup #1 and #2.

| Sample | CS[1)] | Normalized to $^{99}Ru/^{101}Ru$ | | | | | Normalized to $^{102}Ru/^{100}Ru$ | | | | |
|---|---|---|---|---|---|---|---|---|---|---|---|
| | | $\varepsilon^{96}Ru$ (± 2σ) | $\varepsilon^{98}Ru$ (± 2σ) | $\varepsilon^{100}Ru$ (± 2σ) | $\varepsilon^{102}Ru$ (± 2σ) | $\varepsilon^{104}Ru$ (± 2σ) | $\varepsilon^{96}Ru$ (± 2σ) | $\varepsilon^{98}Ru$ (± 2σ) | $\varepsilon^{99}Ru$ (± 2σ) | $\varepsilon^{101}Ru$ (± 2σ) | $\varepsilon^{104}Ru$ (± 2σ) |
| ***Bridgewater_1*** | #1 | 0.25 ± 0.10 | 0.49 ± 0.10 | -0.97 ± 0.04 | -0.44 ± 0.05 | | 2.35 ± 0.12 | 2.06 ± 0.11 | 1.25 ± 0.06 | 0.71 ± 0.03 | |
| ***Bridgewater_2*** | #1 | 0.18 ± 0.11 | 0.04 ± 0.11 | -1.02 ± 0.04 | -0.44 ± 0.05 | | 2.40 ± 0.12 | 1.71 ± 0.12 | 1.31 ± 0.06 | 0.75 ± 0.03 | |
| ***Bridgewater_3*** | #1 | 0.04 ± 0.11 | 0.24 ± 0.12 | -1.02 ± 0.04 | -0.36 ± 0.06 | | 2.44 ± 0.11 | 1.87 ± 0.13 | 1.34 ± 0.06 | 0.70 ± 0.03 | |
| ***Bridgewater_4*** | #1 | 0.07 ± 0.10 | 0.14 ± 0.11 | -1.03 ± 0.04 | -0.42 ± 0.05 | | 2.42 ± 0.13 | 1.79 ± 0.16 | 1.34 ± 0.06 | 0.72 ± 0.04 | |
| ***Bridgewater_5*** | #2 | 0.09 ± 0.08 | 0.10 ± 0.09 | -1.02 ± 0.03 | -0.44 ± 0.04 | -0.04 ± 0.07 | 2.32 ± 0.09 | 1.76 ± 0.09 | 1.31 ± 0.04 | 0.73 ± 0.03 | -0.20 ± 0.04 |
| ***Bridgewater_6*** | #2 | 0.12 ± 0.08 | 0.12 ± 0.08 | -0.96 ± 0.03 | -0.40 ± 0.04 | -0.09 ± 0.08 | 2.25 ± 0.09 | 1.68 ± 0.09 | 1.24 ± 0.05 | 0.69 ± 0.03 | -0.23 ± 0.04 |

Uncertainties represent internal errors of the measurement, i.e., 2 SE.

[1)] Denotes the cone setup used.

Table S4. Ruthenium isotope data used for calculation of the BSE Ru isotope composition.

| Sample | Ref. | Normalized to $^{99}Ru/^{101}Ru$ | | | | | Normalized to $^{102}Ru/^{100}Ru$ | | | | |
|---|---|---|---|---|---|---|---|---|---|---|---|
| | | $\varepsilon^{96}Ru$ (± 2σ) | $\varepsilon^{98}Ru$ (± 2σ) | $\varepsilon^{100}Ru$ (± 2σ) | $\varepsilon^{102}Ru$ (± 2σ) | $\varepsilon^{104}Ru$ (± 2σ) | $\varepsilon^{96}Ru$ (± 2σ) | $\varepsilon^{98}Ru$ (± 2σ) | $\varepsilon^{99}Ru$ (± 2σ) | $\varepsilon^{101}Ru$ (± 2σ) | $\varepsilon^{104}Ru$ (± 2σ) |
| $C3_{mean}$ | 1 | 0.25 ± 0.15 | 0.32 ± 0.42 | 0.00 ± 0.01 | 0.04 ± 0.04 | 0.13 ± 0.18 | 0.34 ± 0.17 | 0.37 ± 0.42 | 0.03 ± 0.02 | -0.02 ± 0.02 | 0.05 ± 0.20 |
| $L5_3_{mean}$ | 1 | 0.29 ± 0.20 | -0.16 ± 0.64 | -0.01 ± 0.03 | 0.03 ± 0.21 | 0.13 ± 0.30 | 0.38 ± 0.48 | -0.12 ± 0.67 | 0.02 ± 0.11 | -0.01 ± 0.11 | 0.05 ± 0.53 |
| $B\text{-}1_{mean}$ | 1 | 0.31 ± 0.11 | 0.15 ± 0.36 | 0.00 ± 0.02 | 0.01 ± 0.08 | 0.15 ± 0.10 | 0.33 ± 0.20 | 0.16 ± 0.37 | 0.00 ± 0.05 | -0.01 ± 0.05 | 0.13 ± 0.20 |
| Kushva mean | 1 | 0.51 ± 0.99 | 0.75 ± 3.28 | 0.03 ± 0.11 | -0.08 ± 0.72 | 0.26 ± 0.37 | 0.26 ± 1.76 | 0.61 ± 3.36 | -0.08 ± 0.38 | 0.03 ± 0.38 | 0.46 ± 1.50 |
| Nizhny $Tagil_{mean}$ | 1 | 0.21 ± 1.25 | 0.03 ± 2.47 | 0.09 ± 0.08 | 0.17 ± 0.72 | 0.52 ± 0.14 | 0.28 ± 1.91 | 0.02 ± 2.58 | -0.05 ± 0.37 | -0.13 ± 0.37 | 0.27 ± 1.45 |
| $Yodda_{mean}$ | 1 | 0.42 ± 0.30 | 0.80 ± 0.14 | 0.04 ± 0.03 | -0.10 ± 0.05 | 0.10 ± 0.20 | 0.10 ± 0.32 | 0.62 ± 0.15 | -0.11 ± 0.04 | 0.03 ± 0.04 | 0.34 ± 0.23 |
| C3 mean | 2 | -0.04 ± 0.11 | 0.01 ± 0.19 | 0.04 ± 0.06 | 0.01 ± 0.09 | -0.05 ± 0.22 | -0.14 ± 0.25 | -0.06 ± 0.23 | -0.06 ± 0.08 | -0.03 ± 0.08 | -0.03 ± 0.31 |
| Bushveld UG2 | 3 | 0.14 ± 0.24 | 0.24 ± 0.49 | 0.05 ± 0.09 | -0.11 ± 0.14 | -0.24 ± 0.22 | -0.23 ± 0.42 | 0.03 ± 0.53 | -0.13 ± 0.12 | 0.03 ± 0.12 | 0.03 ± 0.41 |
| Mean Bushveld | 4 | -0.13 ± 0.07 | -0.01 ± 0.13 | 0.00 ± 0.02 | 0.00 ± 0.02 | 0.02 ± 0.04 | -0.13 ± 0.09 | -0.01 ± 0.14 | 0 ± 0.02 | 0 ± 0.02 | 0.02 ± 0.07 |
| BSE mean | | 0.22 ± 0.16 | 0.24 ± 0.26 | 0.03 ± 0.03 | 0.00 ± 0.07 | 0.11 ± 0.16 | 0.13 ± 0.18 | 0.18 ± 0.22 | -0.04 ± 0.04 | -0.01 ± 0.04 | 0.15 ± 0.13 |

References: 1) Bermingham and Walker (2017) (values rounded); 2) Fischer-Gödde et al., (2015); 3) Fischer-Gödde and Kleine (2017); 4) Fischer-Gödde et al., (2020).

Data from Bermingham and Walker (2017) originally reportet in μ-notation were converted in ε-noation.

All data were originally reported using the $^{99}Ru/^{101}Ru$-normalization and were converted to $^{102}Ru/^{100}Ru$-normalization (see main text for details).

Uncertainties of the BSE mean represent the 95% confidence intervals of the mean (i.e., (s.d. ×t0.95, N-1)/√N).

**References**

Bermingham, K.R., Tornabene, H.A., Walker, R.J., Godfrey, L.V., Meyer, B.S., Piccoli, P., Mojzsis, S.J., 2025. The non-carbonaceous nature of Earth's late-stage accretion. Geochim. Cosmochim. Acta 392, 38-51.

Budde, G., Tissot, F.L., Kleine, T., Marquez, R.T., 2023. Spurious molybdenum isotope anomalies resulting from non-exponential mass fractionation. Geochemistry 83, 126007.

Dauphas, N., Schauble, E.A., 2016. Mass fractionation laws, mass-independent effects, and isotopic anomalies. Annual Review of Earth and Planetary Sciences, 44, 709-783.

Fischer-Gödde, M., Elfers, B.M., Münker, C., Szilas, K., Maier, W.D., Messling, N., Morishita, T., Van Kranendonk, M., Smithies, H., 2020. Ruthenium isotope vestige of Earth's pre-late-veneer mantle preserved in Archaean rocks. Nature 579(7798), 240-244.

Fitoussi, C., Pili, E., Yobregat, E., Touboul, M., Gardin, C., 2025. Identification of mass-independent Mo isotope anomalies in natural and industrially processed samples. Proof of concept for uranium provenance in nuclear forensics. Earth Planet. Sci. Lett., 666, 119459.

Hopp, T., Budde, G., Kleine, T., 2020. Heterogeneous accretion of Earth inferred from Mo-Ru isotope systematics. Earth Planet. Sci. Lett., 534, 116065.

Messling, N., Willbold, M., Kallas, L., Elliott, T., Fitton, J. G., Müller, T., and Geist, D., 2025. Ru and W isotope systematics in ocean island basalts reveals core leakage. Nature, 642, 376-380.

Vance, D., Thirlwall, M., 2002. An assessment of mass discrimination in MC-ICPMS using Nd isotopes. Chem. Geol., 185, 227-240.

Worsham, E.A., Kleine, T., 2021. Late accretionary history of Earth and Moon preserved in lunar impactites. Science advances 7(44).

Young, E.D., Galy, A., Nagahara, H., 2002. Kinetic and equilibrium mass-dependent isotope fractionation laws in nature and their geochemical and cosmochemical significance. Geochim. Cosmochim. Acta, 66, 1095-1104.

Yu, Y., Hathorne, E., Siebert, C., Gutjahr, M., Fietzke, J., Frank, M., 2024. Unravelling instrumental mass fractionation of MC-ICP-MS using neodymium isotopes. Chem. Geol., 662, 122220.